\documentclass[sigconf, screen]{acmart} 
\AtBeginDocument{%
  \providecommand\BibTeX{{%
    \normalfont B\kern-0.5em{\scshape i\kern-0.25em b}\kern-0.8em\TeX}}}

\setcopyright{acmlicensed}
\copyrightyear{2024}
\acmYear{2024}
\acmDOI{10.1145/3613904.3642104}

\acmConference[CHI '24]{Proceedings of the CHI
Conference on Human Factors in Computing Systems}{May 11--16,
2024}{Honolulu, HI, USA}
\acmBooktitle{Proceedings of the CHI Conference on Human Factors in
Computing Systems (CHI '24), May 11--16, 2024, Honolulu, HI, USA} 
\acmISBN{979-8-4007-0330-0/24/05}

\usepackage{float}
\usepackage{hyperref}
\usepackage{balance}
\usepackage{lscape}

\usepackage[shortlabels]{enumitem}

\usepackage{xspace} 

\usepackage{multirow}
\usepackage{fontawesome5}

\usepackage{pifont}
\let\oldding\ding
\renewcommand{\ding}[2][1]{\scalebox{#1}{\oldding{#2}}}

\newcommand{\sig}[1]{\textcolor{purple}{\textbf{#1}}}

\newcommand{\varname}[1]{\uppercase{\texttt{#1}}}

\usepackage{soul}       

\newenvironment{quot}[2]{
\begin{quote}
\begin{center}
\textit{``#1''}
\end{center}

\begin{flushright}
\textit{#2}
\end{flushright}
\end{quote}
}

\usepackage{xcolor,colortbl}
\definecolor{Title-blue}{HTML}{30479c}
\definecolor{SubTitle-blue}{HTML}{6a80d1}
\definecolor{Investor-blue}{HTML}{c7d5ee}
\definecolor{Non-investor-dark-red}{HTML}{c46156}
\definecolor{Non-investor-red}{HTML}{eacdbe}
\definecolor{All-gray}{HTML}{dddcdc}
\definecolor{palered}{HTML}{ffadad}
\definecolor{paleorange}{HTML}{ffd6a5}
\definecolor{paleyellow}{HTML}{fdffb6}
\definecolor{palegreen}{HTML}{caffbf}
\definecolor{palecyan}{HTML}{9bf6ff}
\definecolor{paleblue}{HTML}{a0c4ff}
\definecolor{palepurple}{HTML}{bdb2ff}
\definecolor{palepink}{HTML}{ffc6ff}
\definecolor{palewhite}{HTML}{fffffc}

\begin{document}
\emergencystretch 2em
\title{Privacy in Immersive Extended Reality: Exploring User Perceptions, Concerns, and Coping Strategies}


\author{Hilda Hadan}
\email{hhadan@uwaterloo.ca}
\orcid{0000-0002-5911-1405}
\affiliation{
    \institution{Stratford School of Interaction Design and Business, University of Waterloo}
    \city{Waterloo}
    \country{Canada}
}

\author{Derrick M. Wang}
\email{dwmaru@uwaterloo.ca}
\orcid{0000-0003-3564-2532}
\affiliation{
    \institution{Stratford School of Interaction Design and Business, University of Waterloo}
    \city{Waterloo}
    \country{Canada}
}

\author{Lennart E. Nacke}
\email{lennart.nacke@acm.org}
\orcid{0000-0003-4290-8829}
\affiliation{
\institution{Stratford School of Interaction Design and Business, University of Waterloo}
\city{Waterloo}
\country{Canada}}

\author{Leah Zhang-Kennedy}
\email{lzhangkennedy@uwaterloo.ca}
\orcid{0000-0002-0756-0022}
\affiliation{
    \institution{Stratford School of Interaction Design and Business, University of Waterloo}
    \city{Waterloo}
    \country{Canada}
}


\renewcommand{\shortauthors}{Hadan, et al.}

\begin{abstract}

Extended Reality (XR) technology is changing online interactions, but its granular data collection sensors may be more invasive to user privacy than web, mobile, and the Internet of Things technologies. Despite an increased interest in studying developers' concerns about XR device privacy, user perceptions have rarely been addressed. We surveyed 464 XR users to assess their awareness, concerns, and coping strategies around XR data in 18 scenarios. Our findings demonstrate that many factors, such as data types and sensitivity, affect users' perceptions of privacy in XR. However, users' limited awareness of XR sensors' granular data collection capabilities, such as involuntary body signals of emotional responses, restricted the range of privacy-protective strategies they used. Our results highlight a need to enhance users' awareness of data privacy threats in XR, design privacy-choice interfaces tailored to XR environments, and develop transparent XR data practices.

\end{abstract}



\received{21 September 2023}
\received[revised]{12 December 2023}
\received[accepted]{19 January 2024}
\maketitle

\section{Introduction}
\label{sec:Introduction}

Extended Reality (XR) is the umbrella term for immersive technologies that includes Virtual Reality (VR), Augmented Reality (AR), and Mixed Reality (MR). XR lets users easily create, collaborate, and traverse immersive digital environments~\cite{XRSI2020Definition}. These environments offer realistic visual and auditory experiences, as well as interactive elements. XR is revolutionizing the way people live, work, and play. Following Meta's introduction of the first commercially available VR equipment to the general public in 2016~\cite{Meta2015First}, the XR market has experienced rapid growth, reaching a market size of \$3.31 billion USD in 2023~\cite{statista2022XR}. Since then, XR technologies have been widely used in various domains such as gaming~\cite{shafer2019factors}, social networking~\cite{maloney2020anonymity,maloney2021stay}, research experiments~\cite{mathismoving}, education~\cite{gulhane2019security,pan2006virtual}, and healthcare~\cite{shin2015effects,haworth2012phovr}.

XR technologies create a virtual world for people to interact with digital objects by tracking their positions, movements, and surroundings through body sensors~\cite{buck2022security}. As XR changes the way people access digital information and interact with the real world, advertisers can profile users using XR sensor data~\cite{mhaidli2021identifying}. The immersive features of XR create unique privacy challenges that cannot be easily mitigated or recognized by users. For example, sensors used to operate XR devices can access behavioural pattern data that is traditionally considered non-sensitive to identify users~\cite{miller2020personal, pfeuffer2019behavioural}. Hyperpersonalized avatars that imitate trusted loved ones may steal personal information from users~\cite{mhaidli2021identifying}. Furthermore, the potential to make inferences about users based on data collected from XR systems~\cite{roesner2014security} and the ambiguity surrounding multi-user access control~\cite{maloney2020anonymity, lebeck2018towards}, are emerging issues that raise crucial privacy concerns. The data collected from the XR systems can be used to deduce personal details about people, including their location, interests, and behaviour. The data can be used for targeted advertising or criminal purposes, infringing on people's privacy rights.   

Existing literature had explored privacy issues around AR browsers \cite{mcpherson2015no}, social VR~\cite{maloney2020anonymity,maloney2021stay,o2016convergence}, VR learning~\cite{gulhane2019security}, and behavioural biometrics~\cite{pfeuffer2019behavioural,miller2020personal} from the developers' perspective. These investigations have employed various methodologies to explore the potential privacy risks posed by XR, such as analyzing the system functional requirements~\cite{mcpherson2015no, o2016convergence,dick2021balancing}, using threat modeling of internal and external vulnerabilities~\cite{gulhane2019security}, and surveying existing privacy policies~\cite{dick2021balancing,adams2018ethics}. However, the perceptions of XR users were rarely assessed. 

The objective of our research is to \textit{understand users' awareness and privacy concerns in XR, and to identify the contributing factors (e.g., data type, device status) in raising or inhibiting their concerns.} For users who express privacy concerns in XR, we also study their coping strategies to protect their data privacy. We use a scenario-based survey method to answer the following Research Questions (RQs):

\begin{enumerate}[leftmargin=0.5in,label=\textbf{RQ\arabic*:}] 
    \item How much are people aware of data collection through XR devices?
    \item To what extent are people concerned about their privacy and data collection in XR?
    \item What factors contribute to user privacy concerns in XR?
    \item What coping strategies (if any) do people use to mitigate their privacy concerns?
\end{enumerate}



Our work makes three main contributions. First, we found that XR users lack awareness that XR sensors can capture highly granular data, including involuntary body signals of emotional responses, which could potentially make XR more invasive to their privacy than other types of devices they use on the web, mobile, and Internet of Things (IoT) devices. Second, our survey identified the impacts of various factors through 18 scenarios, such as data type and data sensitivity, on XR users' privacy concerns. Although several of these privacy-related factors were also observed in non-XR environments in previous research (e.g.,~\cite{naeini2017privacy,lee2016understanding, lin2012expectation,tsai2009s,leon2013matters}), our work provides a more precise understanding of the effects of the factors on users in XR environments. Third, we show that participants lack sophisticated mitigation strategies to address their privacy concerns. For example, despite expressing discomfort with data collection, most of our participants opted to ``give up'' protecting their privacy.
Qualitative feedback from participants indicates that they were very pessimistic about their privacy due to the belief that they had already lost privacy on other non-XR devices they use. Furthermore, their limited privacy-protective strategies are likely based on their experience interacting with non-XR devices. Therefore, we suggest that understanding and improving users' mental models of data privacy threats in XR is vital in supporting effective XR risk communication. To achieve transparent XR data practices, the design of privacy-choice interfaces tailored to XR environments is necessary, such as taking into consideration their immersive nature in design to minimize user disruption.

\section{Background and Related Work}
\label{sec:Literature_Review}
Extended Reality (XR) is an umbrella term that describes the following immersive technologies~\cite{XRSI2020Definition}. 

\textit{Virtual Reality (VR):}
VR technologies bring people into completely immersive virtual reality, according to the XR Safety Initiative (XRSI)~\cite{XRSI2020Definition}. This software-generated three-dimensional (3D) environment substitutes reality with a virtual world by covering the user's eyes with equipment like head-mounted displays (HMDs) (e.g., Oculus Rift~\cite{Meta2015First}). VR provides an immersive view that separates the user from the real environment. It enables the user to virtually visit remote locations far from their actual location~\cite{speicher2019mixed}.  

\textit{Augmented Reality (AR):}
Unlike VR, AR refers to technologies that augment, rather than replace, the real world~\cite{XRSI2020Definition}. AR technology overlays digital visuals on top of real-world items to provide users with a more engaging experience. For example, the mobile game \textit{Pokémon GO}\footnote{Pokémon GO official website.~\url{https://pokemongolive.com/}} uses augmented reality to display computer-generated monsters on lawns and sidewalks as players stroll through their areas. AR is user-dependent and always takes place in the physical area surrounding users~\cite{speicher2019mixed}.

\textit{Mixed Reality (MR):}
MR technologies exist at the confluence of VR and AR, mixing the virtual environment with the users' real-world~\cite{XRSI2020Definition,speicher2019mixed}. The virtual and physical worlds coexist and interact in MR. Virtual objects behave and function much like physical items in the real world. For example, the illumination of the virtual object may match the physical light source, or the virtual objects may sound as though they are physically separated from the users. Although most MR systems currently rely on visual and acoustic feedback, researchers are working on technologies that will give haptics, taste, smell, and other cues in the future to simulate temperature, balance, and other factors~\cite{speicher2019mixed}.  

Given that XR encompasses VR, AR, and MR, any devices such as HMDs, smart glasses, controllers, and projectors are considered XR~\cite{XRSI2020Definition}. These devices rely on similar sensors to gather information about the user and their environment to produce the corresponding visual, audio, and haptic effects to simulate an immersive and interactive user experience. Given the similarity in embedded sensors and data tracking capabilities across VR, AR, and MR, we do not focus on a specific XR system. Instead, our study aims to investigate user privacy concerns raised by XR sensors that are generalizable to all three XR systems. 

\subsection{Privacy Risks in Extended Reality}
\label{subsec:Privacy_in_XR}

XR uses body sensors to track user movements and generate visual, audio, and haptic feedback when interacting with the virtual environment~\cite{buck2022security}. Based on these data, inferences can be made about user's physical and mental conditions~\cite{o2016convergence,miller2020personal} as well as habitual movements~\cite{miller2020personal,pfeuffer2019behavioural, maloney2021social}. Cognitive, emotional, and personality issues can be estimated~\cite{buck2022security, mhaidli2021identifying, o2016convergence,o2023privacy}. To better use XR systems, trade-offs between providing users' biometric and demographic information to enable system functionality are inevitable~\cite{maloney2020anonymity}. All this information can further be used to deanonymize the user's identity~\cite{miller2020personal, pfeuffer2019behavioural, gugenheimer2022novel}, manipulate the user toward certain behaviours and buying decisions~\cite{mhaidli2021identifying,buck2022security,o2016convergence}, and derive profit for the manufacturer and third-party companies~\cite{mhaidli2021identifying,egliston2021examining}. Furthermore, some ``always-on'' XR devices enable constant surveillance without user knowledge~\cite{roesner2014security,adams2018ethics,o2023privacy}. Often, users are unaware of the data collection and use~\cite{abraham2022implications,o2023privacy,gallardo2023speculative}. As a result, they might lose control over the information they did not intend to reveal~\cite{o2016convergence}. Additionally, eavesdropping and data breaches could also endanger data privacy~\cite{gulhane2019security}. 

Concerns regarding data tracking sensors and user privacy also arise in multi-user XR environments. While XR enhances communication, it also brings forth privacy risks. Within social VR, participants find greater comfort in discussing their emotions, personal information, and experiences because they believe they remain anonymous or engage with familiar individuals~\cite{maloney2020anonymity}. However, users face the possibility of being deanonymized~\cite{miller2020personal,pfeuffer2019behavioural,gugenheimer2022novel} and deceived by digital avatars impersonating their loved ones or trusted friends~\cite{buck2022security}. Furthermore, because users are always in ecosystems with other users and bystanders, they may accidentally share private information without realizing that others can see it~\cite{lebeck2018towards}. Thus, more personal data becomes open to misuse~\cite{o2016convergence}.  

Furthermore, many studies identified the potential erosion of privacy through bystanders in XR. As unwilling participants in the immersive experiences of others, bystanders might face problems such as being unaware of data collection~\cite{o2023privacy}, being recorded without consent~\cite{o2016convergence}, being recognized unexpectedly by XR applications~\cite{lebeck2018towards,o2023privacy}, and being manipulated virtually (i.e., altering their visual appearance or placing unwanted virtual objects on top of their head) without their authorization~\cite{gugenheimer2022novel,lebeck2018towards,rixen2021exploring}.   




\subsection{Factors Influencing User Privacy Concerns}
\label{subsec:Factors_influencing_privacy_concerns}


Although previous research has investigated potential privacy risks in XR technologies, most of them focused on technical aspects of the issues through threat modelling~\cite{yamakami2020privacy,gulhane2019security}, system functional requirements analyses~\cite{mcpherson2015no, o2016convergence,dick2021balancing}, privacy policy examinations~\cite{dick2021balancing,adams2018ethics}, and expert discussions~\cite{abraham2022implications}. 

Only a few studies focused on privacy concerns from users' perspectives. Among these, many used participants' willingness to share information (measured by the \textit{comfort of data collection}) as an indicator of privacy concerns~\cite{maloney2020anonymity,naeini2017privacy,tsai2009s,lin2012expectation}. For instance, ~\citet{maloney2020anonymity} measured user privacy concerns based on their willingness to share information. They found that social VR users feel comfortable sharing personal information when perceiving anonymity and high familiarity with friends in the virtual environment. ~\citet{adams2018ethics} revealed that VR users' privacy concerns often originate from ``always on'' sensors and the low reputation of device manufacturers. ~\citet{harborth2021investigating} suggested that AR users' privacy concerns are directly driven by perceived permission sensitivity (i.e., perceived sensitivity of the information to which they give permissions), trust in the application, and the general feeling of being a victim of privacy invasion online. Similarly, ~\citet{Coca2019OBrien} revealed that AR users might be comfortable with data collection depending on perceived data sensitivity and the purpose of use. \citet{gallardo2023speculative} found that AR users expressed discomfort about what data were collected and who received the data.

Beyond XR, previous studies have identified various factors that could influence participants' privacy concerns in IoT, mobile, and on the web. For instance, ~\citet{naeini2017privacy} demonstrated that users' comfort with data collection through IoT devices is significantly impacted by the data type, data collection location, and the purpose of data use. ~\citet{lee2016understanding} found that data collections that occurred in personal space, captured photo and voice data, by unknown entities or the government, are considered unacceptable. \citet{hadan2020understanding} found that data collections that are irrelevant to the device's primary functionality raised concerns. Regarding privacy concerns in mobile devices and applications, ~\citet{tsai2009s} indicated that users were more willing to share data when they were informed about data recipients and when the data was not shared. ~\citet{lin2012expectation} revealed that clarifying the purpose of data use reduces the perceived uncertainty of mobile applications and thus increases users' comfort with data sharing. Other studies~\cite{barua2013viewing,klasnja2009exploring,leon2013matters} identified that privacy concerns varied by perceived ownership of data, data retention time, and perceived value of data. In our study, we consider these influential factors from prior research in IoT, mobile, and web contexts to examine their potential impact on user privacy concerns in XR.

Unlike IoT and mobile devices, XR sensors are capable of collecting more granular user data such as eyeball movements~\cite{o2016convergence,egliston2021examining,pfeuffer2019behavioural}, pupil dilation~\cite{pfeuffer2019behavioural,dick2021balancing}, and brain activity~\cite{dick2021balancing,maloney2021social,gugenheimer2022novel} that enabled more privacy-invasive user surveillance and inferences, and raised the discussion around mental privacy~\cite{gugenheimer2022novel} or even neuro-rights~\cite{yuste2021s}. However, most people do not understand how involuntary bodily indicators of emotional responses, mental state, or health can disclose fundamentally private information, such as truthfulness, inner feelings, and sexual arousal~\cite{heller2020watching}. Thus, we see the need for investigating user data collection awareness and privacy concerns in XR.  

\subsection{Research Gap and Connection to our study}
\label{subsec:connection_to_our_study}

Although previous work has explored privacy issues in XR environments, little is known about privacy concerns from the perspectives of XR users. In addition, user-centred XR privacy studies have primarily focused on the requirements of specific XR systems (e.g., AR~\cite{o2023privacy,harborth2021investigating, Coca2019OBrien}, VR~\cite{maloney2020anonymity, adams2018ethics,gallardo2023speculative}), making it necessary to investigate other devices that fall on the Reality-Virtuality Spectrum. Our research lays the groundwork and helps to understand the baseline of user privacy concerns and factors contributing to their comfort with data collection in XR environments. Given the similarity in XR sensors and data collection capabilities, we do not focus on specific XR systems, but aim to investigate users' privacy concerns generalizable across various XR systems. 

We reviewed previous studies on user privacy concerns in XR~\cite{adams2018ethics,roesner2014security,Coca2019OBrien,o2023privacy,gallardo2023speculative}, IoT~\cite{naeini2017privacy,lee2016understanding}, mobile~\cite{lin2012expectation,tsai2009s}, and on the web~\cite{leon2013matters} and iteratively incorporated factors that impacted user privacy perception and preference in our study. We aim to determine whether these factors remained influential on user privacy concerns across various XR systems, and we validate experts' assumptions of privacy issues within XR from users' perspectives~\cite{abraham2022implications}. Given that privacy norms are context-dependent~\cite{nissenbaum2004privacy}, we used a scenario-based approach inspired by prior work~\cite{naeini2017privacy,lee2016understanding,harborth2021investigating}.
As the first attempt to seek essential insights into users' privacy concerns in XR, we narrow our scope by focusing on personal XR systems in single-user scenarios.

\section{Methodology}
\label{sec:methodology}
To understand XR users' privacy concerns, we conducted a scenario-based online survey with 464 Prolific\footnote{Prolific.~\url{https://www.prolific.co/}} participants who had prior experience with XR technologies, but ownership of XR devices was not a requirement. Our study received Research Ethics Board approval (REB\#44772) from the University of Waterloo before recruiting participants. We presented each participant with four XR data collection scenarios, randomly selected from a total of 18 pre-designed scenarios. The randomized selection allowed us to explore a wider range of scenarios without overwhelming participants with a long survey.

Privacy norms are dependent on context~\cite{nissenbaum2004privacy}. Scenarios are hypothetical situations in specific circumstances~\cite{finch1987vignette} that have been frequently used in prior studies to provide participants with contexts when evaluating privacy concerns~\cite{naeini2017privacy,harborth2021investigating}. For this reason, we developed scenarios to support our investigation of user privacy concerns in XR. We incorporated factors that were found to influence user privacy perception and preference in XR~\cite{adams2018ethics,roesner2014security,Coca2019OBrien}, IoT~\cite{naeini2017privacy,lee2016understanding}, mobile~\cite{lin2012expectation,tsai2009s}, and on the web~\cite{leon2013matters}. Our aimed to determine whether these factors remained influential in XR environments.

In the following sections, we describe the design of the scenarios, the development of the survey questions, and our participant recruitment process. 

\subsection{Scenarios Development}
\label{subsec:scenarios_Design}

Drawing from previous studies on factors that potentially influence users' privacy perception in XR~\cite{adams2018ethics,roesner2014security,Coca2019OBrien,o2023privacy,gallardo2023speculative}, IoT~\cite{naeini2017privacy,lee2016understanding}, mobile~\cite{lin2012expectation,tsai2009s}, and on the web~\cite{leon2013matters}, and discussions with our research team of privacy experts, we iteratively identified five high-level factors that are most likely to affect people's privacy concerns in XR technologies.
Three are static factors (i.e., \varname{user\_type}, \varname{location} and \varname{device\_type}) that remain unchanged throughout all scenarios, and two are dynamic factors (i.e., \varname{data\_type} and \varname{device\_status}) that varied between scenarios. 

\textit{Static Factors:}
\begin{enumerate}
        \item \textbf{User Type} (\varname{user\_type}): the type of user being involved in data collection; Always set to the user (i.e., ``you'')
        \item \textbf{Location} (\varname{location}): the location where the data collection occurs; Always set to a private space (i.e., ``home / personal room'')
        \item \textbf{Device Type} (\varname{device\_type}): the type of technology that collects data; Always set to XR devices that includes VR, AR, and MR (i.e., ``XR devices'') 
    \end{enumerate}

\textit{Dynamic Factors:}
\begin{enumerate}[resume]
        \item \textbf{Data Type} (\varname{data\_type}): The type of data collected by XR devices;
        \item \textbf{Device Status} (\varname{device\_status}): Whether the data collection happens actively (i.e., when the user is interacting with the device) or passively (i.e., when the device is running in the background). 
    \end{enumerate}

Two researchers with expertise in usable privacy reviewed and discussed these factors in multiple iterations, and developed more granular levels for each of the five factors identified above. We illustrate our process below. 

\subsubsection{User Type, Location, and Device Type (factors \#1–3)}
Our aims to understand the baseline of user privacy concerns and factors that contribute to their comfort with data collection across various XR environments. While previous studies found privacy issues from both users' and bystanders' perspectives~\cite{gugenheimer2022novel,o2016convergence,lebeck2018towards,o2023privacy} and in multi-user environments~\cite{rixen2021exploring,maloney2020anonymity}, as discussed in~\autoref{subsec:connection_to_our_study}, we narrow our scope by focusing on personal XR systems in single-user scenarios. Thus, our data collection \varname{location} was ``Home/ personal room'' for all scenarios, and the \varname{user\_type} was always ``You'', representing the user themselves.
Furthermore, our study described data collection \varname{device\_type} as ``XR devices'' as a whole and did not differentiate between specific XR systems in our scenarios. The decision was made to keep the number of scenarios manageable for the participants and to take into consideration that, in most cases, VR, AR, and MR sensors share similar data collection sensors. Although some privacy concerns, such as bystander effects, may be more likely to occur in AR, users with virtual avatar representations may become bystanders for other users in VR~\cite{denning2014situ}, and disconnecting from physical surroundings can cause VR users to be monitored by bystanders in the physical world without their knowledge~\cite{do2023vice}. Therefore, by treating XR systems as a cohesive unit, our study focuses on shared privacy concerns raised by common XR sensors and data practices, rather than focusing on the device type and manufacturer.




\subsubsection{Device Status (factor \#4)}

Previous research found that user privacy concerns could originate from their interaction with XR devices and devices with continuously operating sensors~\cite{adams2018ethics}. Thus, our factor \varname{device\_status} encompasses two levels: \textit{in-use} representing situations when the user actively interacts with the device, and \textit{background-running} representing situations when the device is running in the background. 

\subsubsection{Data Type (factor \#5)}

Several studies have tried to classify the data types that XR technology collects~\cite{XRSI2021Virtual,dick2021balancing,o2023privacy,gallardo2023speculative}. The primary framework we chose for scenario design is a XR data classification from an article by~\citet{dick2021balancing} because it clearly described each category of XR data, taking user privacy into account. Based on the framework, we incorporated data types previously examined in XR studies to ensure the inclusion of a diverse range of data types in our scenarios. 

We initially compiled 26 data types that XR sensors could capture based on the framework from \citet{dick2021balancing}. These data types comprise user account information (e.g., demographics, billing address, phone number)~\cite{adams2018ethics,dick2021balancing,o2023privacy}, device information (e.g., crash report, model information, system logs)~\cite{dick2021balancing}, user geo-location~\cite{o2016convergence,gulhane2019security,mcpherson2015no}, user audio data~\cite{adams2018ethics,egliston2021examining,mcpherson2015no}, infrared camera data or user images~\cite{adams2018ethics,egliston2021examining,o2016convergence,o2023privacy}, user movements~\cite{de2019security,miller2020personal,maloney2021social,o2023privacy}, user visual attention~\cite{de2019safemr,pfeuffer2019behavioural,egliston2021examining}, physical body dimensions~\cite{adams2018ethics,egliston2021examining,o2023privacy}, user surroundings~\cite{egliston2021examining,dick2021balancing,mcpherson2015no,o2023privacy}, and digital communication messages and friend list. 
Since our research aims to investigate XR-related concerns, we further refined the list of data types by excluding common data types such as user geo-location, facial features, payment information, and IP address that were already investigated in previous mobile, web, and IoT privacy studies. Thus, in our scenarios, we focused only on nine data types that are frequently collected by XR devices, but are less prevalent in other technologies. To enhance participants' comprehension of each data type, descriptive examples enclosed in ``()'' are also presented to them in the scenarios. 

Overall, our \varname{data\_type} factor includes nine levels in four data categories~\cite{dick2021balancing}:
\begin{enumerate}[label=\alph*]
    \item \textit{Observable data:} data about an individual that third-parties can observe and replicate, which includes:
    \begin{itemize}
    \item Physical appearance (e.g., body dimensions)~\cite{adams2018ethics,egliston2021examining,o2016convergence,o2023privacy}; 
    \item Identifying in-app/in-world assets (e.g., personal virtual objects, personal avatars)~\cite{dick2021balancing,o2023privacy};
    \item Environment information and dimensions (e.g., users' surroundings, room layout, device/user position in relation to the environment, user position in relation to device)~\cite{egliston2021examining,dick2021balancing,mcpherson2015no,o2023privacy}.
    \end{itemize}
    \item \textit{Observed data:} data that individual provides or generates, which third-parties cannot replicate, including:
    \begin{itemize}
        \item Physical movements and characteristics (e.g., head/hand/ body/eye motion, orientation, position, gestures, posture, fitness information)~\cite{dick2021balancing,buck2022security,gugenheimer2022novel,de2019security,miller2020personal,maloney2021social,lebeck2018towards,XRSI2020Definition,o2023privacy};
        \item Physiological data (e.g., pupil and cornea reflections, brainwaves, skin signals)~\cite{buck2022security,gugenheimer2022novel,lebeck2018towards,XRSI2020Definition,o2023privacy}.
    \end{itemize}
    \item \textit{Computed data:} new data inferred by manipulating observable and observed data, including:
    \begin{itemize}
        \item Visual attention (e.g., eye gaze, area of interest, fixation, heatmaps, time to first fixation, time spent on a certain point, fixation sequences)~\cite{buck2022security,de2019safemr,pfeuffer2019behavioural,egliston2021examining,XRSI2020Definition,o2023privacy};
        \item Mannerisms (e.g., gait, habitual movements)~\cite{dick2021balancing,miller2020personal,pfeuffer2019behavioural, maloney2021social};
        \item Cognitive, emotional, and personality estimates (e.g., cognitive load, stress, depression, excitement, identity traits, etc.)~\cite{buck2022security, mhaidli2021identifying, o2016convergence, miller2020personal,o2023privacy}
    \end{itemize}
    \item \textit{Associated data:} data that, on its own, does not provide descriptive details about an individual, such as:
    \begin{itemize}
        \item Non-identifying virtual assets (e.g., in-app achievements)~\cite{dick2021balancing}.
    \end{itemize}
\end{enumerate}



Finally, the combination of three static factors and two dynamic factors with nine data type categories and two device statuses resulted in a total of 18 scenarios shown in~\autoref{tab:Scenarios}, developed using a scenario template: 

\begin{quot}{\ul{You}\textit{ are at }\ul{home/personal room}\textit{, and your }\ul{XR device}\textit{ (e.g., smart headset, touch controller, 3D projector) is keeping track of your} [\varname{data\_type}] \textit{when} [\varname{device\_status}].}{}
\end{quot}

\begin{table*}[!t]
\centering
\caption{Summary of the 18 scenarios used in the study. ID= Scenario identifier, $n$ = number of participants presented with each scenario. Each scenario was presented to approximately 103 participants on average.}
\label{tab:Scenarios}
\resizebox{\textwidth}{!}{%
\begin{tabular}{@{}lll@{}}
\toprule
\textbf{ID} & \multirow{3}{*}{$n$} & \textbf{Scenarios} \\
& & \textit{\underline{All scenarios begin with}}: \\
\textit{Scenario:Type-Status} &  & ``You are at home/personal room, and your XR device (e.g., smart headset, touch controller, 3D projectors) is keeping track of your...'' \\ \midrule
S:Appearance-Use & 99 & Physical appearance (e.g., body dimensions) when you use the device. \\
S:Appearance-Bg & 107 & Physical appearance (e.g., body dimensions) when the device is running in the background. \\
S:Identifying-Use & 105 & Identifying in-app/in-world assets (e.g., personal virtual objects, personal avatars) when you use the device. \\
S:Identifying-Bg & 103 & Identifying in-app/in-world assets (e.g., personal virtual objects, personal avatars) when the device is running in the background. \\
S:Environment-Use & 102 & \multirow{2}{*}{\begin{tabular}[c]{@{}l@{}}Environment information and dimensions (e.g., users' surroundings, room layout, device/user position in relation to the environment,\\user position in relation to device) when you use the device.\end{tabular}} \\
 &  &  \\
S:Environment-Bg & 107 & \multirow{2}{*}{\begin{tabular}[c]{@{}l@{}}Environment information and dimensions (e.g., users' surroundings, room layout, device/user position in relation to the environment, \\user position in relation to device) when the device is running in the background.\end{tabular}} \\
 &  &  \\
S:Movements-Use & 106 & \multirow{2}{*}{\begin{tabular}[c]{@{}l@{}}Physical movements and characteristics (e.g., head/hand/body/eye motion, orientation, position, gestures, posture, fitness information)\\when you use the device. \end{tabular}}\\
 &  &  \\
S:Movements-Bg & 103 & \multirow{2}{*}{\begin{tabular}[c]{@{}l@{}}Physical movements and characteristics   (e.g., head/hand/body/eye motion, orientation, position, gestures, posture, fitness information)\\when the device is running in the background.\end{tabular}} \\
 &  &  \\
S:Physiological-Use & 104 & Physiological data (e.g., pupil and cornea reflections, brainwaves, skin signals) when you use the device. \\
S:Physiological-Bg & 102 & Physiological data (e.g., pupil and   cornea reflections, brainwaves, skin signals) when the device is running in the background. \\
S:Visual-Use & 99 & \multirow{2}{*}{\begin{tabular}[c]{@{}l@{}}Visual attention (e.g., eye gaze, area of interest, fixation, heatmaps, time to the first fixation, time spent on a certain point, fixation\\sequences) when you use the device.\end{tabular}} \\
 &  &  \\
S:Visual-Bg & 101 & \multirow{2}{*}{\begin{tabular}[c]{@{}l@{}}Visual attention (e.g., eye gaze, area of interest, fixation, heatmaps, time to the first fixation, time spent on a certain point, fixation \\sequences) when the device is running in the background.\end{tabular}} \\
 &  &  \\
S:Mannerisms-Use & 100 & Mannerisms (e.g., gait, habitual movements) when you use the device. \\
S:Mannerisms-Bg & 106 & Mannerisms (e.g., gait, habitual movements) when the device is running in the background. \\
S:Cognitive-Use & 104 & Cognitive, emotional, and personality estimates (e.g., cognitive load, stress, depression, excitement, identity traits, etc.) when you use\\
& & the device. \\
S:Cognitive-Bg & 102 & \multirow{2}{*}{\begin{tabular}[c]{@{}l@{}}Cognitive, emotional, and personality   estimates (e.g., cognitive load, stress, depression, excitement, identity traits, etc.) when the device\\ is running in the background.\end{tabular}} \\
 &  &  \\
S:Non-identifying-Use & 102 & Non-identifying virtual assets (e.g., in-app achievements) when you use the device. \\
S:Non-identifying-Bg & 107 & Non-identifying virtual assets (e.g., in-app achievements) when the device is running in the background. \\ \bottomrule
\end{tabular}%
}
\end{table*}

\vspace{-5mm}
\subsection{Survey Design}
\label{subsec:Survey_Design}


\subsubsection{Screening Questionnaire.} The survey began with a study information sheet and consent form, followed by a screening questionnaire. We set our screening criteria to recruit XR users from North America and Europe, the leading regions in the XR market~\cite{Fortune2021Extended}. The North American region has been a market leader in early XR device adoption, with many well-known vendors, and the European Commission actively supports research and innovation in XR technologies. To participate, individuals had to be at least 18 years old and possess some familiarity using XR devices, but ownership of XR devices was not mandatory. 

\subsubsection{XR Understanding.}
To ensure that participants' privacy concerns are specifically related to their experience in XR environments, we first evaluated participants' understanding of VR, AR, and MR technologies. In a drag and drop pair-matching question, participants were asked to match the different descriptions of XR technologies to the corresponding definitions. 
After this exercise, we provided participants with the correct descriptions of VR, AR, and MR definitions from the XR Safety Initiative (XRSI)~\cite{XRSI2020Definition} to ensure that have an appropriate baseline understanding of different types XR technologies to answer the remaining questions in the survey.

\subsubsection{Perceptions and Comfort Towards Data Collection.} 
To assess participants' awareness of XR's data collection capability, they were asked to share what data they thought were being observed by XR devices based on their experience. 
Then, the participants were presented with four randomly selected scenarios from 18 scenarios (see \autoref{tab:Scenarios}). For each scenario, we adapt the methodology of~\citet{naeini2017privacy}, where participants were asked to rate how likely the scenario is to happen today, within 2 years, and within 10 years, to assess how \textit{realistic} participants think the scenarios are, with the scenarios perceived to happen today being the most realistic. We coded their responses as \varname{likely\_today}, \varname{likely\_in\_2yrs}, and \varname{likely\_in\_10yrs}. These questions were answered on 5-point Likert scales from \textit{1-extremely unlikely} to \textit{5-extremely likely}.





To prevent biasing participants' perceptions, we did not explicitly mention ``privacy'' or related terms when measuring their privacy concerns. Instead, we followed the approach used in previous studies (see \autoref{subsec:Factors_influencing_privacy_concerns}) to measure participants' comfort level with data collection (\varname{comfort\_level}) 
using a 5-point Likert-scale from \textit{1-very uncomfortable} to \textit{5-very comfortable}). Participants were also asked to rate the perceived sensitivity of the data type (\varname{data\_sensitivity}) on a 5-point Likert-scale from \textit{1-not sensitive} to \textit{5-very sensitive}). For participants who expressed discomfort with the situation described in the scenario, we asked them to share their mitigation strategy in an open-ended question.   

\subsubsection{Experience and Privacy Concern.} Finally, we inquired about participants' personal experience with XR devices, including how many years they have been using XR devices (\varname{exp\_years}), the type of devices used (\varname{exp\_device}), the frequency of use (\varname{exp\_frequency}), and the purpose for using these devices (\varname{exp\_purpose}). We measured participants' general level of privacy concern on the Internet using the Internet Users' Information Privacy Concerns Scale (IUIPC)~\cite{malhotra2004internet}, which consists of three dimensions of user privacy concerns: collection (\varname{IUIPC\_collection}), control (\varname{IUIPC\_control}), and awareness (\varname{IUIPC\_awareness}). 
To ensure data quality, we included an attention check question mixed in the scenarios.
We closed the survey with demographic questions. \autoref{fig:survey-flow} summarizes the survey flow. A complete set of questions is included in the Supplementary Materials. 

\begin{figure*}[!t]
\centering
  \includegraphics[width=\textwidth]{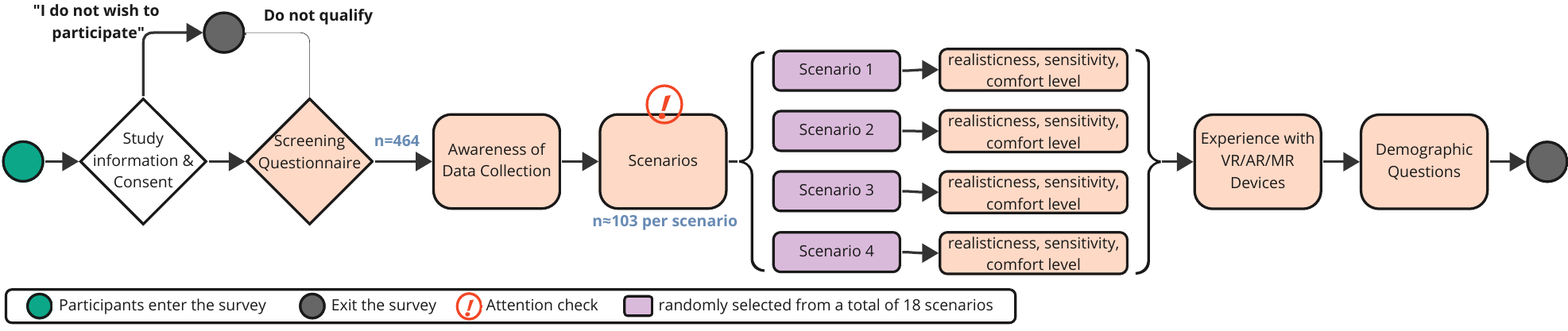}
  \vspace{-7mm}
  \caption{Survey flow: Upon passing the screening questionnaire, 464 participants were asked about their understanding of XR and awareness of data collection. Next, they were presented with four randomly selected scenarios from a total of 18 scenarios from \autoref{tab:Scenarios}, resulting in approximately 103 responses per scenario. For each scenario, participants answered questions relating to comfort with data collection, perceived data sensitivity, and how realistic they perceived the scenario to be. Finally, the participants shared their experience with XR devices and demographic information.}
  \label{fig:survey-flow}
\end{figure*}

\subsection{Participants Recruitment and Demographics}
\label{subsec:Participants_Recruitment}

We piloted the survey questionnaire with 15 students. Using their feedback, we improved the questions' wording and order to enhance understandability. 
A power analysis~\cite{faul2009statistical} for a Wilcoxon-signed rank test determined that we need a minimum of $57$ participants for each pair of scenarios to achieve a power of $\geq95\%$, allowing a margin for random error of $\leq5\%$ with an estimated effect size=0.50.
We deployed our survey on Prolific in December 2022 and received $549$ responses with an average completion time of 13 minutes. We rejected 49 incomplete responses. The remaining 500 participants were remunerated US~\$5.8 each\footnote{US~\$5.8 is about CAD~\$8. This remuneration rate is calculated based on Ontario local minimum hourly wage. See:~\url{https://www.immigrationwaterlooregion.ca/en/study-and-work/salary-standards-and-minimum-wage.aspx}}. We excluded 23 responses that failed the attention check question and another 13 responses that were automatically identified as bots by Qualtrics. 
Our final analysis was based on the remaining $N=464$ valid responses, with $n\approx64$ for each pair of scenarios, providing sufficient data for assessing significant within-subjects differences on participants' privacy concerns (measured as \varname{comfort\_level}).


\begin{table*}[!ht]
\centering
\caption{Breakdown of our participants' ($N=464$) gender, age, education level, income level, and three IUIPC dimensions. A detailed overview of participants' demographic is located in \autoref{tab:demographic} in Supplementary Materials.}
\label{tab:short-demographic}
\resizebox{\textwidth}{!}{%
\begin{tabular}{@{}llllllllll@{}}
\toprule
\multicolumn{2}{c}{\textbf{Gender}} & \multicolumn{2}{c}{\textbf{Age}} & \multicolumn{2}{c}{\textbf{Education Level}} & \multicolumn{2}{c}{\textbf{Income}} & \multicolumn{2}{c}{\textbf{IUIPC factors}} \\ \midrule
Man & \multicolumn{1}{l|}{230 (50\%)} & Range & \multicolumn{1}{l|}{18-56} & No high school & \multicolumn{1}{l|}{4 (<1\%)} & Less than \$25,000 & \multicolumn{1}{l|}{182 (39\%)} & \multicolumn{2}{l}{\textit{\underline{IUIPC\_Collection}}} \\
Woman & \multicolumn{1}{l|}{222 (48\%)} & Mean (SD) & \multicolumn{1}{l|}{27.78 (7.95)} & High school & \multicolumn{1}{l|}{173 (37\%)} & \$25,000-\$49,999 & \multicolumn{1}{l|}{151 (33\%)} & Range & (1.25 - 7) \\
Non-binary & \multicolumn{1}{l|}{11 (2\%)} & &  \multicolumn{1}{l|}{} & Associates & \multicolumn{1}{l|}{30 (7\%)} & \$50,000-\$99,999 & \multicolumn{1}{l|}{86 (19\%)} & Mean (SD) & 5.69 (1.14) \\
Prefer not to say & \multicolumn{1}{l|}{1 (<1\%)} &  & \multicolumn{1}{l|}{} & Bachelors & \multicolumn{1}{l|}{162 (35\%)} & \$100,000-\$199,999 & \multicolumn{1}{l|}{19 (4\%)} & \underline{\textit{IUIPC\_Control}} &  \\
 & \multicolumn{1}{l|}{} &  & \multicolumn{1}{l|}{} & Professional & \multicolumn{1}{l|}{94 (20\%)} & More than \$200,000 & \multicolumn{1}{l|}{5 (1\%)} & Range & (3 - 7) \\
 & \multicolumn{1}{l|}{} &  & \multicolumn{1}{l|}{} & No answer & \multicolumn{1}{l|}{1 (<1\%)} & Prefer not to say & \multicolumn{1}{l|}{21 (5\%)} & Mean (SD) & 5.78 (0.92) \\
 & \multicolumn{1}{l|}{} &  & \multicolumn{1}{l|}{} &  & \multicolumn{1}{l|}{} &  & \multicolumn{1}{l|}{} & \multicolumn{2}{l}{\underline{\textit{IUIPC\_Awareness}}} \\
 & \multicolumn{1}{l|}{} &  & \multicolumn{1}{l|}{} &  & \multicolumn{1}{l|}{} &  & \multicolumn{1}{l|}{} & Range & (3 - 7) \\
 & \multicolumn{1}{l|}{} &  & \multicolumn{1}{l|}{} &  & \multicolumn{1}{l|}{} &  & \multicolumn{1}{l|}{} & Mean (SD) & 6.29 (0.74) \\ \bottomrule
\end{tabular}%
}
\end{table*}

We summarize participants' demographic background in \autoref{tab:short-demographic}.
Our 464 participants consist of $50\%$ men, $48\%$ women, $2\%$ non-binary/third gender, and less than $1\%$ chose not to disclose their gender. The majority of our participants, aged between 20 and 29, possessed an Associate degree or higher (Associate degree $=6\%$, Bachelor's degree $=34\%$, Graduate or professional degree $=20\%$), with an annual income of less than \$50,000 USD (Less than \$25,000 $=39\%$, \$25,000 to \$49,999 $=32\%$) and displayed relevantly high IUIPC scores (see \autoref{tab:short-demographic}). The participants came from 21 countries in North America and Europe (see \autoref{tab:demographic} in Supplementary Materials). 

The majority of the participants ($94\%$) used XR devices for entertainment purposes. Regarding their experience with XR devices, most of the participants had less than five years of experience using XR devices (Less than 6 months $=40\%$; 6 months to 1 year $=26\%$; 1 to 5 years $=36\%$). In terms of frequency, a small number of participants engaged with XR devices at least once a day ($5\%$), while others used them once a week ($25\%$) or once a month ($26\%$). When it comes to specific types of XR devices, the majority of participants ($85\%$) had experience with smart headsets or glasses, and most ($58\%$) had experience with handheld devices (see \autoref{tab:demographic} in Supplementary Materials). 



\subsection{Data Analysis}
\label{subsec:Data_Analysis}

We demonstrate the details of our quantitative data analysis while introducing our results in \autoref{sec:Findings}. For the open-ended question,
we analyzed the responses using inductive thematic analysis with two researchers~\cite{braun_thematic_2012}. We started by familiarizing ourselves by scanning through the data and excluding blank or incoherent responses. We kept responses such as ``none'' or ``nothing'' because having no way to mitigate privacy concerns was still a valid response (see numbers of included responses for each question in~\autoref{fig:discomfort-open-ended}). The two researchers first coded approximately 10\% of the data individually, then discussed and resolved conflicts. This process was repeated twice until the codebook was finalized and all remaining data were coded. Using the codebook, the researchers developed and refined the themes. A summary of themes and codes is shown on~\autoref{tab:codebook}.

\vspace{-1mm}
\section{Findings}
\label{sec:Findings}

\subsection{RQ1: How much are people aware of the data collected through XR devices?}
\label{subsec:findings-RQ1}

\begin{figure*}[!ht]
\centering
  \includegraphics[width=0.7\textwidth]{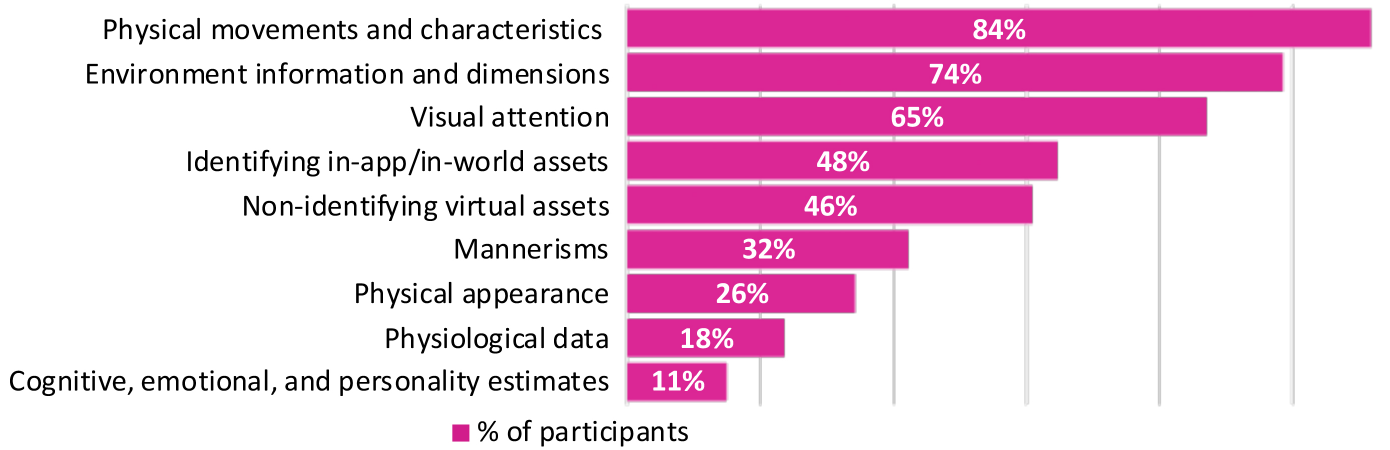}
  \vspace{-3mm}
  \caption{Percentage of participants' ($N=464$) perceived data collection of various data types through XR devices.} 
  \label{fig:awareness}
\end{figure*}

\autoref{fig:awareness} illustrates the overall response distribution among the $N=464$ participants regarding the types of data they believed XR devices could capture. The majority of participants ($84\%$) acknowledged that XR devices can collect \textit{Physical movements and characteristics}, while $74\%$ recognized the capture of \textit{Environment information and dimensions} through XR devices. Furthermore, a large proportion of participants realized that their \textit{Visual attention} ($65\%$), \textit{Identifying in-app/in-world assets} ($48\%$), and \textit{Non-identifying virtual assets} ($46\%$) are getting observed by the XR devices. Some participants also believed that XR devices observe their \textit{Mannerisms} ($32\%$), \textit{Physical appearance} ($26\%$), and \textit{Physiological data} ($18\%$). Only a small number of participants ($11\%$) thought that their \textit{Cognitive, emotional, and personality estimates} could be observed. A few participants also mentioned \textit{Other} information, such as ``purchases and utilization of assets'' ($\leq1\%$), while a small percentage ($1\%$) were uncertain about what data their devices might observe. 

\subsection{RQ2: To what extend are people concerned about their privacy regarding data collection in XR?}
\label{subsec:findings-RQ2}

\begin{figure*}[!ht]
\centering
  \includegraphics[width=0.85\textwidth]{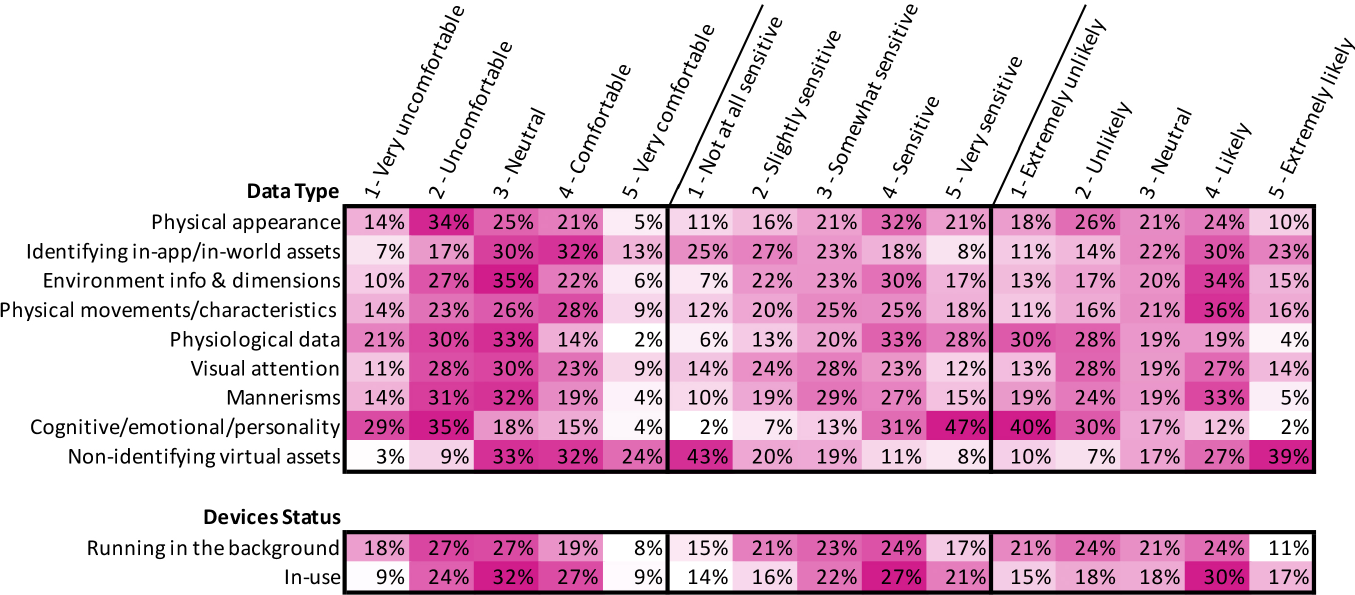}
  \vspace{-2mm}
  \caption{Percentage of participants' ($n=464$) comfort level, perceived data sensitivity and perceived realism of a scenario between the data type and the device status. Cells with a higher percentage have a darker background colour. For instance, 29\% of participants selected very uncomfortable when their \textit{cognitive, emotional, and personality estimates} were being observed by the device.}
  \label{fig:factors-all}
\end{figure*}

The distribution of participants' \varname{comfort\_level} across various data types and device statuses are shown in \autoref{fig:factors-all}. To validate the observed differences in participants' comfort level, we performed a set of pairwise Wilcoxon-signed-rank comparisons~\cite{woolson2007wilcoxon} with Bonferroni correction~\cite{chen2017general} (see~\autoref{tab:wilcoxon-signed-rank-data-type} and~\autoref{tab:wilcoxon-signed-rank-device-status} in Supplementary Materials). 

Among all scenarios, participants expressed relevantly strong discomfort when the data type associated with the scenario was \textit{Cognitive, emotional, and personality estimates}, with 29\% of participants selected very uncomfortable and 35\% selected uncomfortable. This difference in participants' perceived comfort level was significant (\textit{P}$\leq$\textit{.007}) across all data types (see~\autoref{tab:wilcoxon-signed-rank-data-type}). Moreover, this particular data type was considered the most sensitive and the most unlikely to happen today, with $47\%$ of participants selecting very sensitive and $40\%$ selecting extremely unlikely (see \autoref{fig:factors-all}). This difference in participants' perceived data sensitivity and realisticness were significant (\textit{P}\textit{<.001}) across all data types as well. Compared to other data types, participants exhibited a significant (\textit{P}$\leq$\textit{.005}) higher level of comfort when the scenario involved \textit{Non-identifying virtual assets}, as only 9\% of participants selected unco`mfortable and 3\% selected very uncomfortable. This data type was also perceived as the least sensitive among the others, with 43\% of participants indicating that it was not sensitive. This difference is also significant (\textit{P<.001}) across all data types. 

Furthermore, a slightly greater number of participants expressed discomfort, when the device operates passively in the background ($18\%$ selected very uncomfortable, and 27\% selected uncomfortable). On the other hand, a slightly greater number of participants perceived higher level of data sensitivity and realism when scenarios involving active device interaction, as 27\% and 21\% of participants selected sensitive and very sensitive, and 30\% and 17\% of participants selected likely and extremely likely (see~\autoref{fig:factors-all}). These differences are also significant (\textit{P}$\leq$\textit{.005}) based on our pairwise analyses (see~\autoref{tab:wilcoxon-signed-rank-device-status}).

\subsection{RQ3: What factors contribute to user privacy concerns in XR?}
\label{subsec:findings-RQ3}

This section summarizes participants' perceptions of how realistic the scenarios are, their general privacy concerns on the Internet, and their experience with and frequency of using XR devices. We further compare the influence of these factors on participants' comfort levels towards data collection in XR.

\subsubsection{Factors impacting participants' comfort level}

To evaluate the relationship between participants' comfort levels and various factors, we employed Cumulative Link Mixed Model (CLMM) regression\footnote{We used the Ordinal R-package (\url{https://cran.r-project.org/web/packages/ordinal/}) for modeling participants' comfort level.}, incorporating a random intercept for each participant as part of our models. CLMM regression proves particularly beneficial for analyzing repeated measures experiments with ordinal dependent variable such as our study, where participants were presented with multiple parallel scenarios~\cite{christensen2019tutorial}.

We first conducted a series of Univariate CLMM regressions. Each regression employed the participants' \varname{comfort\_level} as the dependent variable (DV) and one of the factors as the predictor. The purpose was to ascertain the existence of relationships and select predictors for our multivariate model. As shown in \autoref{tab:CLMM-univariate-ordinal}, the results indicated significant relationships between participants' \varname{comfort\_level} (DV) and predictors such as \varname{data\_type} (\textit{P<.001}), \varname{device\_status} (\textit{P<.001}),  perceived \varname{data\_sensitivity} (\textit{P<.001}), \varname{likely\_today}/\varname{in\_2yrs}/\varname{in\_10yrs} (\textit{P<.001}), general privacy concerns on the Internet (IUIPC scores \textit{P<.001}). No significance was found regarding participants' demographic information or previous XR experience.

We further conducted a multivariate CLMM regression using participants' \varname{comfort\_level} as the dependent variable (DV), with significant predictors identified in the univariate CLMM regressions (see \autoref{tab:CLMM-univariate-ordinal}). To obtain the best-fit model, we utilized the backwards elimination method, which involved starting with a full model containing all variables and sequentially eliminating the variable with the highest p-value in each step until reaching the global minimum Akaike information criterion (AIC)~\cite{kadane2004methods}. The final multivariate model, presented in \autoref{tab:CLMM-multivariate-ordinal}, was derived from this process. We arranged the predictors in descending order based on their contribution to the \varname{comfort\_level}, as determined by the AIC values obtained in the univariate CLMM (where each model contained only one predictor and the random intercept). Predictors were ranked with the highest contribution (lowest AIC in univariate CLMM) appearing first. Positive estimates (or OR>1) indicate a tendency towards comfort, while negative estimates (or OR<1) indicate a tendency towards discomfort. 

\begin{table*}[!ht]
\caption{Univariate Cumulative Linked Mixed Model analyses of factors impacting participants' comfort level (5-point Likert data, from 1-very uncomfortable to 5-very comfortable), with a random intercept per participant. Ordinal data are used as is. }
\centering
\resizebox{0.65\textwidth}{!}{%
\begin{tabular}{@{}lllllll@{}}
\toprule
\textbf{Predictor} &  &  &  &  & &\textbf{AIC}\\ \midrule
\textit{\textbf{data\_type}} & \textbf{Estimates} & \textbf{Std. Error} & \textbf{z} &\textbf{\textit{P} value} & \textbf{OR (95\%CI)} & 5185.92\\ \cmidrule(l){2-6} 
Non-identifying virtual assets & Reference &  &  &  &  &  \\
\rowcolor{palered!50}Identifying in-app/in-world assets & -0.84 & 0.20 & -4.13 & <.001\sig{***} & 0.43 (0.29, 0.64) &  \\
\rowcolor{palered!50}Visual attention & -1.67 & 0.21 & -8.05 & <.001\sig{***} & 0.19 (0.13, 0.28) &  \\
\rowcolor{palered!50}Physical movements and   characteristics & -1.45 & 0.20 & -7.14 & <.001\sig{***} & 0.23 (0.16, 0.35) &  \\
\rowcolor{palered!50}Mannerisms & -2.09 & 0.21 & -10.10 & <.001\sig{***} & 0.12 (0.08, 0.19) &  \\
\rowcolor{palered!50}Environment information and   dimensions & -1.72 & 0.20 & -8.57 & <.001\sig{***} & 0.18 (0.12, 0.27) &  \\
\rowcolor{palered!50}Physical appearance & -2.10 & 0.21 & -10.13 & <.001\sig{***} & 0.12 (0.08, 0.18) &  \\
\rowcolor{palered!50}Physiological data & -2.54 & 0.21 & -12.08 & <.001\sig{***} & 0.08 (0.05, 0.12) &  \\
\rowcolor{palered!50} Cognitive, emotional, and   personality estimates & -3.17 & 0.22 & -14.41 & <.001\sig{***} & 0.04 (0.03, 0.07) &  \\ \bottomrule
\textbf{\textit{device\_status}} & \textbf{Estimates} & \textbf{Std. Error} & \textbf{z} &\textbf{\textit{P} value} & \textbf{OR (95\%CI)} & 5442.24 \\ \cmidrule(l){2-6} 
In-use & Reference &  &  &  &  &  \\ 
\rowcolor{palered!50}Running in the background & -0.55 & 0.09 & -5.94 & <.001\sig{***} & 0.58 (0.48, 0.69) & \\ \bottomrule
\textbf{\textit{data\_sensitivity}} & \textbf{Estimates} & \textbf{Std. Error} & \textbf{z} &\textbf{\textit{P} value} & \textbf{OR (95\%CI)} & 4442.67 \\ \cmidrule(l){2-6} 
1 - Not sensitive & Reference &  &  &  &  &  \\
\rowcolor{palered!50}2 & -2.25 & 0.20 & -11.09 & <.001\sig{***} & 0.11 (0.07, 0.16) &  \\
\rowcolor{palered!50}3 & -3.57 & 0.22 & -16.52 & <.001\sig{***} & 0.03 (0.02, 0.04) &  \\
\rowcolor{palered!50}4 & -5.12 & 0.24 & -21.64 & <.001\sig{***} & 0.01 (0.00, 0.01) &  \\
\rowcolor{palered!50}5 - Very sensitive & -6.74 & 0.27 & -24.72 & <.001\sig{***} & 0.00 (0.00, 0.00) & \\ \bottomrule
\textbf{\textit{likely\_today}} & \textbf{Estimates} & \textbf{Std. Error} & \textbf{z} &\textbf{\textit{P} value} & \textbf{OR (95\%CI)} & 4984.37 \\ \cmidrule(l){2-6} 
1 - Extremely unlikely & Reference &  &  &  &  &  \\
\rowcolor{palered!50}2 - Unlikely & 0.40 & 0.18 & 2.26 & .02\sig{*} & 1.49 (1.05, 2.11) &  \\
\rowcolor{palered!50}3 - Neutral & 1.21 & 0.19 & 6.46 & <.001\sig{***} & 3.34 (2.32, 4.82) &  \\
\rowcolor{palered!50}4 - Likely & 2.22 & 0.19 & 11.85 & <.001\sig{***} & 9.17 (6.36, 13.23) &  \\
\rowcolor{palered!50}5 - Extremely likely & 4.12 & 0.23 & 17.61 & <.001\sig{***} & 61.44 (38.85, 97.15) & \\ \bottomrule
\textbf{\textit{likely\_in\_2yrs}} & \textbf{Estimates} & \textbf{Std. Error} & \textbf{z} &\textbf{\textit{P} value} & \textbf{OR (95\%CI)} & 5097.70 \\ \cmidrule(l){2-6} 
1 - Extremely unlikely & Reference &  &  &  &  &  \\
2 - Unlikely & 0.61 & 0.34 & 1.81 & .07 & 1.84 (0.95, 3.58) &  \\
\rowcolor{palered!50}3 - Neutral & 1.21 & 0.34 & 3.59 & <.001\sig{***} & 3.37 (1.73, 6.53) &  \\
\rowcolor{palered!50}4 - Likely & 2.37 & 0.33 & 7.18 & <.001\sig{***} & 10.68 (5.59, 20.40) &  \\
\rowcolor{palered!50}5 - Extremely likely & 3.75 & 0.34 & 10.94 & <.001\sig{***} & 42.31 (21.62, 82.73) & \\ \bottomrule
\textbf{\textit{likely\_in\_10yrs}} & \textbf{Estimates} & \textbf{Std. Error} & \textbf{z} &\textbf{\textit{P} value} & \textbf{OR (95\%CI)} & 5305.95 \\ \cmidrule(l){2-6} 
1 - Extremely unlikely & Reference &  &  &  &  &  \\
2 - Unlikely & -0.26 & 0.51 & -0.50 & 0.62 & 0.77 (0.28, 2.11) &  \\
3 - Neutral & 0.58 & 0.49 & 1.18 & 0.24 & 1.78 (0.69, 4.62) &  \\
\rowcolor{palered!50}4 - Likely & 1.10 & 0.47 & 2.35 & .02\sig{*} & 3.01 (1.20, 7.54) &  \\
\rowcolor{palered!50}5 - Extremely likely & 2.31 & 0.47 & 4.91 & <.001\sig{***} & 10.04 (4.00, 25.22) &  \\ \bottomrule
\textbf{\textit{IUIPC scores}} & \textbf{Estimates} & \textbf{Std. Error} & \textbf{z} &\textbf{\textit{P} value} & \textbf{OR (95\%CI)} & 5409.70 \\ \cmidrule(l){2-6}
\rowcolor{palered!50}IUIPC\_CONTROL & -0.34 & 0.08 & -4.45 & <.001\sig{***} & 0.71 (0.61, 0.83) &   \\
\rowcolor{palered!50}IUIPC\_AWARENESS & -0.36 & 0.10 & -3.80 & <.001\sig{***} & 0.70 (0.58, 0.84) &   \\
\rowcolor{palered!50}IUIPC\_COLLECTION & -0.50 & 0.06 & -8.37 & <.001\sig{***} & 0.61 (0.54, 0.68) &  \\ \bottomrule
\textbf{\textit{Age}} & \textbf{Estimates} & \textbf{Std. Error} & \textbf{z} &\textbf{\textit{P} value} & \textbf{OR (95\%CI)} & 5477.09 \\ \cmidrule(l){2-6}
 & 0.01 & 0.01 & 0.85 & 0.40 & 1.01 (1.00, 1.03) & \\ \bottomrule
\textbf{\textit{Education}} & \textbf{Estimates} & \textbf{Std. Error} & \textbf{z} &\textbf{\textit{P} value} & \textbf{OR (95\%CI)} & 5479.58 \\ \cmidrule(l){2-6}
Less than high school & Reference &  &  &  &  &  \\
High school & 1.89 & 0.82 & 2.32 & 0.12 & 6.65 (1.34, 32.94) &  \\
Associate's degree & 2.11 & 0.86 & 2.46 & 0.28 & 8.26 (1.53, 44.40) &  \\
Bachelor's degree & 1.68 & 0.82 & 2.06 & 0.12 & 5.38 (1.08, 26.65) &  \\
Graduate's degree& 1.77 & 0.82 & 2.15 & 0.25 & 5.86 (1.17, 29.42) &  \\ \bottomrule
\textbf{\textit{exp\_years}} & \textbf{Estimates} & \textbf{Std. Error} & \textbf{z} &\textbf{\textit{P} value} & \textbf{OR (95\%CI)} & 5479.80 \\ \cmidrule(l){2-6}
Less than 6 months & Reference &  &  &  &  &  \\
6 months to 1 year & 0.37 & 0.19 & 1.96 & 0.05 & 1.45 (1.00, 2.11) &  \\
1 year to 5 years & 0.09 & 0.17 & 0.53 & 0.60 & 1.09 (0.79, 1.51) &  \\
5 to 10 years & 0.25 & 0.54 & 0.46 & 0.65 & 1.28 (0.45, 3.67) &  \\
More than 10 years & 0.29 & 1.11 & 0.26 & 0.80 & 1.33 (0.15, 11.75) & \\ \bottomrule
\textbf{\textit{exp\_frequency}} & \textbf{Estimates} & \textbf{Std. Error} & \textbf{z} &\textbf{\textit{P} value} & \textbf{OR (95\%CI)} & 5470.92 \\ \cmidrule(l){2-6}
Once a year & Reference &  &  &  &  &  \\
At least once every six months & -0.25 & 0.27 & -0.92 & 0.36 & 0.78 (0.46, 1.32) &  \\
At least once every three   months & 0.27 & 0.26 & 1.04 & 0.30 & 1.31 (0.79, 2.16) &  \\
At least once a month & 0.13 & 0.23 & 0.57 & 0.57 & 1.14 (0.73, 1.78) &  \\
At least once a week & 0.26 & 0.23 & 1.15 & 0.25 & 1.30 (0.83, 2.03) &  \\
At least once a day & 1.12 & 0.36 & 3.08 & 0.07 & 3.06 (1.50, 6.23) & \\ \bottomrule
\multicolumn{7}{l}{\begin{tabular}[c]{@{}l@{}} \textit{Note. Significance are displayed as: \sig{***} P<.001, \sig{**} P<.01, \sig{*} P<.05. OR=Odds Ratio, CI=Confidence Interval. The Reference}\\ \textit{categories were selected to enhance result interpretability. For OR, a value greater than 1 indicates a positive relationship, and}\\ \textit{a value less than 1 indicates a negative relationship.}\end{tabular}} \\ 
\end{tabular}%
}
\label{tab:CLMM-univariate-ordinal}
\vspace{-2mm}
\end{table*}
\newpage


\begin{table*}[!ht]
\caption{Multivariate Cumulative Linked Mixed Model analyses of factors impacting participants' comfort level (1-very uncomfortable to 5-very comfortable), with a random intercept per participant. Ordinal data are used as is.}
\centering
\resizebox{0.69\textwidth}{!}{%
\begin{tabular}{@{}llllll@{}}
\toprule
\textbf{Predictor} & \textbf{Estimates} & \textbf{Std. Error} & \textbf{z} &\textbf{\textit{P} value} & \textbf{OR (95\%CI)} \\ \midrule
\textbf{\textit{data\_sensitivity}} &  &  &  &  &  \\
1 - Not sensitive & Reference &  &  &  &  \\
\rowcolor{palered!50}2 & -1.83 & 0.21 & -8.66 & <.001\sig{***} & 0.16 (0.11, 0.24) \\
\rowcolor{palered!50}3 & -2.92 & 0.23 & -12.93 & <.001\sig{***} & 0.05 (0.04, 0.08) \\
\rowcolor{palered!50}4 & -4.27 & 0.25 & -17.13 & <.001\sig{***} & 0.01 (0.01, 0.02) \\
\rowcolor{palered!50}5 - Very sensitive & -5.66 & 0.29 & -19.54 & <.001\sig{***} & 0.00 (0.00, 0.01) \\
\textbf{\textit{likely\_today}} &  &  &  &  &  \\
1 - Extremely unlikely & Reference &  &  &  &  \\
2 - Unlikely & 0.29 & 0.19 & 1.54 & .13 & 1.33 (0.92, 1.92) \\
\rowcolor{palered!50}3 - Neutral & 0.81 & 0.20 & 4.09 & <.001\sig{***} & 2.25 (1.53, 3.33) \\
\rowcolor{palered!50}4 - Likely & 1.44 & 0.20 & 7.22 & <.001\sig{***} & 4.22 (2.85, 6.23) \\
\rowcolor{palered!50}5 - Extremely likely & 2.52 & 0.25 & 10.10 & <.001\sig{***} & 12.44 (7.63, 20.31) \\
\textbf{\textit{data\_type}} &  &  &  &  &  \\
Non-identifying virtual assets & Reference &  &  &  &  \\
Identifying in-app/in-world assets & -0.25 & 0.22 & -1.17 & .24 & 0.78 (0.51, 1.19) \\
\rowcolor{palered!50}Visual attention & -0.61 & 0.23 & -2.71 & .007\sig{**} & 0.54 (0.35, 0.84) \\
Physical movements and   characteristics & -0.25 & 0.22 & -1.10 & .27 & 0.78 (0.51, 1.21) \\
\rowcolor{palered!50}Mannerisms & -0.72 & 0.23 & -3.17 & .002\sig{**} & 0.49 (0.31, 0.76) \\
Environment information and dimensions & -0.38 & 0.22 & -1.71 & .09 & 0.69 (0.45, 1.06) \\
\rowcolor{palered!50}Physical appearance & -0.55 & 0.23 & -2.42 & .02\sig{*} & 0.58 (0.37, 0.90) \\
\rowcolor{palered!50}Physiological data & -0.56 & 0.23 & -2.38 & .02\sig{*} & 0.57 (0.36, 0.91) \\
Cognitive, emotional, and personality estimates & -0.47 & 0.25 & -1.85 & .06 & 0.63 (0.38, 1.03) \\
\textbf{\textit{IUIPC scores}} &  &  &  &  &  \\
\rowcolor{palered!50}IUIPC\_COLLECTION & -0.38 & 0.09 & -4.26 & <.001\sig{***} & 0.68 (0.57, 0.82) \\
IUIPC\_CONTROL & -0.22 & 0.11 & -2.02 & .06 & 0.80 (0.65, 0.99) \\
\textbf{\textit{device\_status}} &  &  &  &  &  \\
In-use & Reference &  &  &  &  \\
\rowcolor{palered!50}Running in the background & -0.40 & 0.10 & -3.86 & <.001\sig{***} & 0.67 (0.55, 0.82) \\ \bottomrule
\multicolumn{6}{l}{\begin{tabular}[c]{@{}l@{}} \textit{Note. final AIC = 4244.43. Significance are displayed as follows: \sig{***} P<.001, \sig{**} P<.01, \sig{*} P<.05. OR=Odds Ratio. } \\ \textit{CI=Confidence Interval. The Reference categories were selected to enhance result interpretability. For OR,}\\\textit{a value greater than 1 indicates a positive relationship, and a value less than 1 indicates a negative relationship.}\end{tabular}} \\ 
\end{tabular}%
}
\label{tab:CLMM-multivariate-ordinal}
\end{table*}

As illustrated in \autoref{tab:CLMM-multivariate-ordinal}, the \varname{data\_sensitivity} described in the Scenarios had the greatest negative effect on participants' \texttt{COMFORT\_\\LEVEL} with data collection (\textit{P<.001}). Participants' \varname{comfort\_level} was positively influenced by the perceived likelihood of happening today (\varname{likely\_today}, \textit{P<.001}), indicating a negative association between perceived data sensitivity and participants' comfort, as well as a positive association between the perceived realisticness of the scenario and their comfort with data collection. 

In addition, not all data types yielded statistically significant results. For instance, compared to Scenarios where \textit{Non-identifying virtual assets} were being observed, participants had lower \texttt{COMFORT\_\\LEVEL} when Scenarios collecting \textit{Visual attention} (\textit{P=.007}), \textit{Mannerisms} (\textit{P=.002}), \textit{Physical appearance} (\textit{P=.02}), and \textit{Physiological data} (\textit{P=.02}). That is, participants were more likely to express discomfort when these data types were being collected. On the other hand, we did not identify significance from Scenarios in which \textit{Identifying in-app/in-world assets}, \textit{Physical movements and characteristics}, and \textit{Environment information and dimensions} were observed. This is in line with our results in~\autoref{fig:factors-all}, where participants demonstrated a relevantly neutral \varname{comfort\_level} when the scenario involved the collection of these three data types, possibly due to their unclear (i.e., evenly distributed) \varname{data\_sensitivity} and their lack of understanding of the associated privacy implication. We found evidence from our qualitative data, where more participants sought detailed information about the data collection and surrounding privacy laws (``comprehend and control what happens to my data'') when the scenarios involved these three data types (see~\autoref{fig:discomfort-open-ended}: \varname{Identifying-Use}, \varname{Identifying-Bg}, \varname{Movements-Use}, \varname{Movements\\-Bg}, \varname{Environment-Use}, and \varname{Environment-Bg}).


Participants expressed a strong discomfort with scenarios involving \textit{cognitive, emotional, and personality estimates} and this particular type of data was also considered the most sensitive. In the participants' qualitative responses, they also mentioned that the collection of these types of data is \textit{``too much intrusion''} compared to other types of data. The discomfort may also be related to the trust issue with data recipients and perceived purpose of data use~\cite{gallardo2023speculative,o2023privacy}, as some participants sought to understand data practices and expressed trust issues with the manufacturer and the device system in their qualitative responses (see~\autoref{fig:discomfort-open-ended}). 
However, we did not find a significant difference from scenarios involving these data types. We believe this because while \textit{cognitive, emotional and personality estimates} are perceived as invasive, scenarios involving the data types were also perceived as the least likely to occur today, providing a possible explanation for the lack of significance in our CLMM.
This suggest that if the participants perceive the scenario is unlikely to happen, it may contribute to downplaying the perceived privacy risk~\cite{naeini2017privacy,gallardo2023speculative}. 

Lastly, we found that participants' \varname{comfort\_level} was negatively affected by their \varname{IUIPC\_collection} score (\textit{P<.001}), suggesting that their concerns about the fairness of data collection (i.e., cost and benefits) was negatively impacting their comfort level. We also identified a negative relationship between participants' \varname{comfort\_level} and the \varname{device\_status} (\textit{P<.001}), suggesting that participants were more likely to express discomfort when the device is running in the background. These results correspond to our qualitative analysis, where many participants expressed concerns about perceived privacy risks, such as fear of surveillance, use of data for other than the stated purpose (misuse), and lack of trust in how sensitive information would be handled by entities collecting or processing it 
(see Section~\ref{subsec:findings-RQ4}). 

\subsection{RQ4: What coping strategies (if any) do people use to mitigate their privacy concerns?}
\label{subsec:findings-RQ4}

\begin{figure*}[!t]
\centering
  \includegraphics[width=\textwidth]{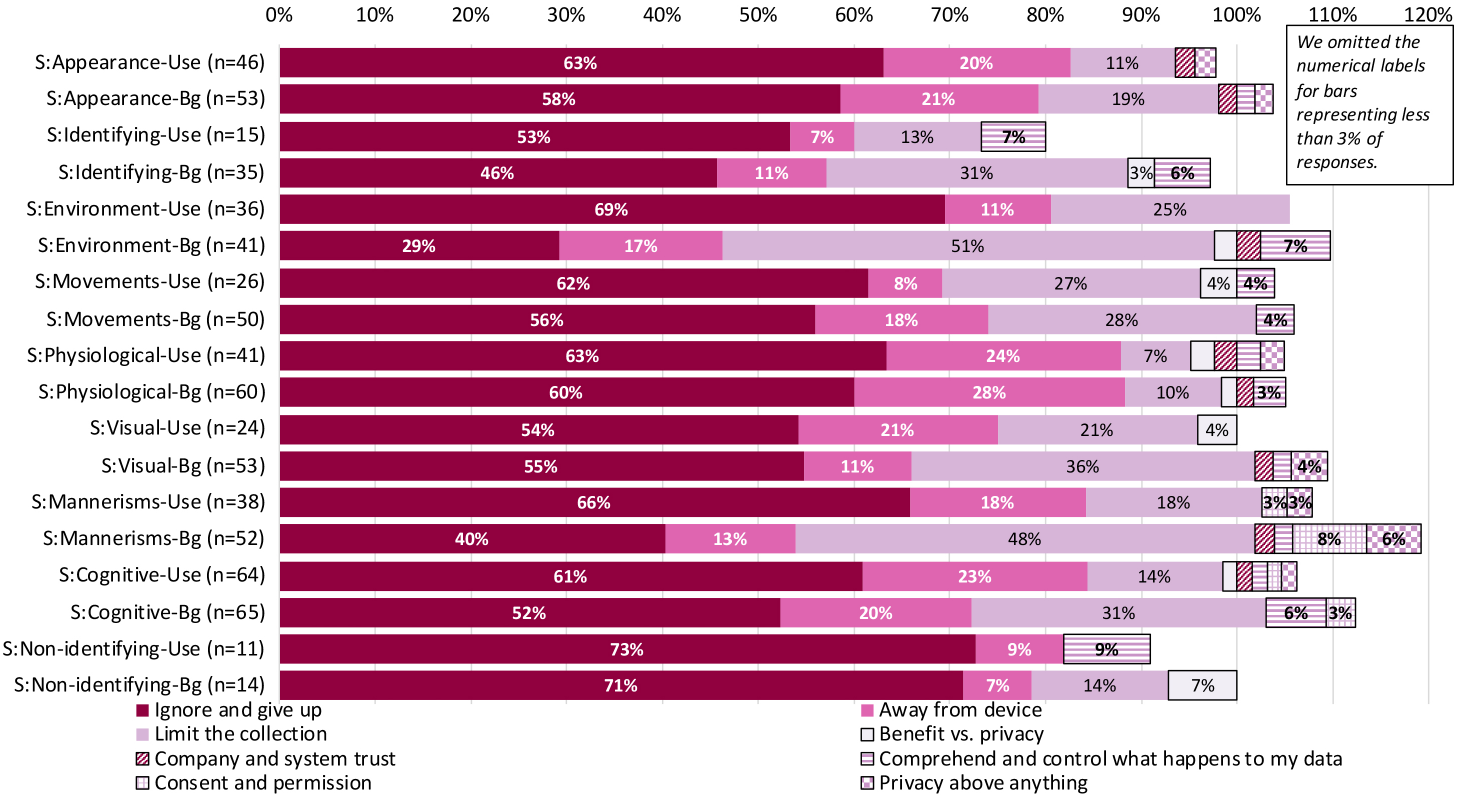}
  \vspace{-5mm}
  \caption{n= number of participants who responded to this question. Summary of participants' responses to the conditional open-ended question ``What would you do (if any) to mitigate the discomfort raised?'' (Q5) for each Scenario. The question was only shown to participants who selected \textit{1-very uncomfortable} or \textit{2-uncomfortable} regarding data collection (Q3) in each Scenario. Total percentage $>100\%$ since a participant might mention multiple coping strategies.} 
  \vspace{-3mm}
  \label{fig:discomfort-open-ended}
\end{figure*}

This section further discusses the themes derived from the participants' qualitative responses (\autoref{fig:discomfort-open-ended}) and how they relate to the corresponding quantitative responses. For clarity, themes are placed within quotations and participants' quotes are set in \textit{italics}. We use a percentage range (e.g., $7\%\sim28\%$) to represent diverse proportions of participants that mentioned a same theme across different scenarios, and we use a single percentage (e.g., $3\%$) for themes mentioned in a specific scenario or by a specific participant group. The questions and themes can be found in the Supplementary Materials (Q5 and~\autoref{tab:codebook}).

Across all 18 Scenarios, a large portion of the participants expressed an inability to mitigate discomfort towards potential data collection and stated their intention to continue using the device (``ignore or give up''). From their qualitative responses, we identified that these participants were pessimistic about their privacy within the XR environment and had become accustomed to the absence of privacy. 

\begin{quot}{Technology will overcome us. I think I would feel strange at first, but eventually I would get used to, even tho I would rather not share this information with unknowned entities. We don't have privacy anymore, I realize that.}{--- Participant 313}
\end{quot}

Notably, for Scenarios that involve the same data type, a higher proportion of participants indicated their willingness to disregard potential data collection when actively using the device. This result aligns with our findings in~\autoref{fig:factors-all}, where a higher number of participants expressed discomfort when the device operated passively in the background. In addition, a significantly lower number of participants expressed discomfort when \textit{Non-identifying virtual assets} were collected in the Scenario ($n=11$ in \varname{S:Non-Identifying-Use}, $n=14$ in \varname{S:Non-Identifying-Bg}). Most of these participants ($73\%$ in \varname{S:Non-Identifying-Use} and $71\%$ in \varname{S:Non-Identifying-Bg}) expressed the tendency to ``ignore and give up''. This finding could be explained by our quantitative results, where \textit{Non-identifying virtual assets} were perceived as the least sensitive. 

While the majority of participants opted to \textit{``give up''}, some simultaneously claimed that nothing can be done \textit{``unless I'm aware of data collection.''} This suggests that raising awareness and transparency about XR data collection might encourage participants to safeguard their privacy in XR. In addition, $29\%$ of participants mentioned about ignoring the discomfort when their \textit{Environment info \& dimensions} were collected by an XR device running in the background in \varname{S:Environment-Bg} because, in contrast to other data types, participants perceived the devices that track environmental data as easily containable. Most of them (51\%) suggested isolating the device by placing it in a designated and non-personal room (\textit{``limit the data collection''}).  


In fact, $7\%\sim51\%$ of participants proposed the idea of \textit{``limiting the data collection''} as the strategy to safeguard their privacy. This encompassed actions like disabling device sensors, placing devices in designated rooms with low privacy concerns, disconnecting or unplugging the device, uninstalling privacy-invasive applications, adjusting privacy settings, or installing specialized applications to intervene in the data collection processes. Some participants engaged in self-censorship behaviours, such as using the device only when necessary, avoiding regular mannerisms, or suppressing genuine emotions to prevent monitoring.

\quot{I would probably change my mannerism and try to limit my displays of any sort of emotions from time to time to feel like my `real me' is not being monitored.}{--- Participant 215}

Moreover, some participants ($7\%\sim28\%$) expressed their intention to stay ``away from the device,'' either by discontinuing its use or disposing of the device. A few participants (3\%) also indicated that they would have refrained from purchasing such devices had they been aware of the data collection beforehand. 
Conversely, six participants (1\%) weighed the device's benefits and the potential compromise to their privacy when deciding to purchase or use it.  

Furthermore, $\sim9\%$ of the participants wanted greater and more effective control over their data.
Some mentioned actively reading user agreements, privacy policies, data privacy laws, and online articles to seek detailed information about the collected data. Others reported exploring options to reduce the amount and type of data being collected and methods to delete their data from the device. They emphasized that the data should be anonymized and only used to benefit users' lives and experiences, and deemed data use for targeted advertising or behaviour manipulation inappropriate.
In addition, $\sim8\%$ of participants emphasized the importance of informed consent. 
In general, participants expressed a desire to understand what is being collected, how it is used, and with whom their data is shared. 

\quot{I don't like the idea of my personal data being collected by other companies for their own profit. Show the users that data is only being used for their interest and not with second intentions.}{--- Participant 163}
Lastly, $\sim18\%$ of the participants prioritized their privacy over all other XR features and stated their intention to switch to alternative models or manufacturers that value and protect their privacy. On the contrary, $\sim2\%$ leaned towards completely trusting the device manufacturer 
and are willing to accept any data-related processes.

\vspace{-1mm}
\section{Discussion}
\label{sec:Discussion}

We first compare our study with privacy studies on other types of devices. We then discuss the implications of our findings to future work, including the importance of addressing users' unawareness of XR data privacy threats and passive XR data collection processes, the need for designing privacy-choice interfaces tailored to XR environments and implementing privacy-friendly default settings to reduce user burden, and the necessity to enforce and develop transparent XR data practices through XR product design and new regulations. 

\subsection{Similarities and Differences Between XR Privacy and Other Types of Devices}

Many of our findings supported the results from prior work in IoT (e.g.,~\cite{naeini2017privacy,lee2016understanding,oulasvirta2012long}), mobile (e.g.,~\cite{lin2012expectation,tsai2009s}), and web privacy studies (e.g.,~\cite{leon2013matters}), suggesting a convergence in users' privacy concerns regardless of settings. For instance, our regression analysis revealed that users' privacy concerns (comfort with data collection) were driven by their perceived data sensitivity (e.g.,~\cite{harborth2021investigating,Coca2019OBrien}), data type (e.g.,~\cite{naeini2017privacy}), perceived equity of data collection (i.e., the IUIPC\_collection~\cite{harborth2021investigating}), and their perceived realism of the scenario (e.g.,~\cite{naeini2017privacy}). Our study also found a significant association between users' privacy concern and device status, aligning with concerns regarding ``always-on'' devices from~\citet{adams2018ethics} and~\citet{roesner2014security}. Although our participants' demographic information and experience with XR devices do not affect their privacy concerns, we suspect it is primarily due to their lower exposure with new commercial XR technologies. 

In addition, our qualitative analysis showed that participants mainly agreed on several privacy-seeking strategies frequently mentioned in non-XR environments. For example, our participants reported self-censorship in response to the presence of nearby XR devices, such as a ``chilling effect'' of being more conscious of their speech and emotions~\cite{lebeck2018towards,o2016convergence}. Their reported strategy of removing the device or blocking sensors echoes previous findings on personal space control challenges in AR~\cite{lebeck2018towards}, VR threats of recording devices in private places~\cite{o2016convergence}, AR bystanders' interests in physical or technical measures~\cite{denning2014situ}, and users' strategies in dealing with IoT surveillance~\cite{oulasvirta2012long}.

Pessimistically, there was a strong agreement among our participants on the difficulty of protecting their privacy in XR. Many participants felt a sense of privacy resignation because they believed that they had already lost their privacy on other devices. On the other hand, stopping using of the device was embraced by a notable group of participants who value personal privacy more than the innovative functionality that XR provides. The underlying core agreement was the assumption that their data is collected to support other commercial activities, such as targeted advertising. This assumption was confirmed by a network traffic analysis on Oculus VR~\cite{trimananda2022ovrseen}. 
While infrequently mentioned, our qualitative responses identified participants' trust issues with the manufacturer and the XR system, and their desire to comprehend XR data use and to give explicit consent on XR data collection, echoing findings in IoT studies~\cite{harborth2021investigating,naeini2017privacy}.



While we found that XR privacy has some of the same concerns as other types of technology, we suggest that interaction paradigms and privacy interfaces users hold in 2D applications will need to be adapted and expanded.

\subsection{Implications for Future Work}


\subsubsection{Data Transparency and Awareness in XR}
Unlike web, mobile, and some IoT devices, XR 
collects users' physical, physiological, environmental, and even emotional estimates to 
produce 
the immersive experience~\cite{cummings2016immersive}. Data are often collected without explicit notice or consent~\cite{trimananda2022ovrseen}. Although our participants were generally comfortable with XR data collection, which was slightly higher than the comfort level with IoT devices in previous studies (e.g.,~\cite{naeini2017privacy}), we believe it is in part due to their limited awareness of the types, uses, and management of their XR data~\cite{o2023privacy,gallardo2023speculative,van2017better}. For example, only a few participants recognized that physiological and physical data could be observed and mannerisms and cognitive, emotional, and personality estimates could be inferred from other data (see~\autoref{fig:awareness}). These data types could serve as a behavioural identifier~\cite{gugenheimer2022novel, miller2020personal, pfeuffer2019behavioural} and could make users vulnerable to privacy risks from emotional manipulations~\cite{mhaidli2021identifying}. The lack of awareness further leads to misconceptions between perceived risks and benefits (i.e., privacy calculus~\cite{dinev2006extended}). 

Since many users are new to XR technologies, it is crucial for product and policymakers to ensure that people are aware of various types of data used by XR technologies in 3D environments that are not generally required for 2D web and mobile-based applications. In fact, since we found the types data that are least understood (e.g., cognitive, emotions, and personality estimates) are not generally used on non-XR devices, we suspect that many users rely on their privacy experiences on non-XR devices to make privacy decisions in XR environments. While users' experience with other types of technology may make them more familiar with some types of data collection---such as location tracking and voice recording---data types like movement tracking may be less understood, particularly when the scale and nature of their use may change in XR technologies to adapt to new use cases.

While some participants showed indifference to XR data collection, many still expressed discomfort towards data sharing in XR---especially regarding passive data collection and inferences made from sensors when a person is not intentionally sharing their data (e.g., emotional inferences from facial expressions and physiological sensor inputs). One possible explanation could be that it is difficult for users to assess data collected from unconscious action and behaviours (e.g., system `reads' users' facial expressions to make product recommendations) because it is invisible to the users and often operating in the background, compared to more active data collection (e.g., users intentionally smile to signal agreement). This is evident by our qualitative data that a slightly higher proportion of participants mentioned to ``comprehend and control what happens to my data'' when the scenario involved a background running device (see~\autoref{fig:discomfort-open-ended}). This suggests that in addition to \textit{what} novel data types are collected, the \textit{way} the data are collected can also influence people's comfort with the data. Therefore, XR designers should help users be aware of the data they provide to the device, especially those data collected 
from unconscious processes when users have little awareness. As XR continues to innovate, we encourage future research to explore users' perceptions on emerging data types and explore visual and creative ways to communicate data collection beyond 2D interfaces. The results from the research could inform XR designers about the types of data that are considered more intrusive and should be minimized.

\subsubsection{Informed Data Collection without Sacrificing Enjoyment and Immersion}
Many participants wanted to be informed when data collection happens. Some mentioned about trying to limit XR device data collection by blocking sensors physically, practicing self-censorship, and discontinuing its use. However, these strategies often meant sacrificing the enjoyment and functionality of their XR devices. One way to achieve informed data collection and allow users to opt-out when necessary is through notifications in XR.
However, creating efficient notifications in XR environments is a challenging task. To date, various forms of XR notifications have been developed to inform users through head-up display, on-body haptic feedback, floating display, and in-situ display on a surrounding virtual object~\cite{rzayev2019notification}. On the one hand, the rich information and feedback that XR provides often leads to disconnection from the real world and causes important notifications being missed~\cite{rzayev2019notification}. On the other hand, notifications that distract users' attention in XR are often found to disrupt the user experience~\cite{george2018intelligent,kan2021popup}. 
Thus, future XR designers and researchers should explore ways to integrate XR notifications that better balance their effectiveness and their disruptiveness to the immersive experience. 
3D interactions facilitated by XR technology are naturally different than those enabled by 2D screen-based interfaces. Therefore, the way users interact with and in 3D environments provides an opportunity for privacy researchers to study how these 3D interactions (e.g., gestural and spatial controls) can help users understand how their data are collected and processed. For example, digital objects can become privacy interfaces and user consent might be communicated through a range of physical gestures and embodied actions that feel more natural in XR environments than text, icons, and menu-based consent notices. As the design space for XR privacy expands, we encourage designers to prioritize functionalities that are less intrusive to user privacy. This requires future research to explore alternative designs that achieve similar functionality but involve less sensitive data types. Our findings revealed participants' perceived data sensitivity and comfort level for a comprehensive list of data types.

Another way to achieve immersion is for XR designers to implement privacy-friendly default settings~\cite{GDPR-privacy-by-default}. For instance, 
default settings can restrict the sharing of user information~\cite{yao2019defending,GDPR-privacy-by-default}. While having a privacy-friendly default can be effective in reducing users' burden in making the `better' privacy decisions by offering a path of least resistance, researchers worry that nudging users with these defaults may threaten their autonomy. A person's decision to take an action or change a setting might become biased outside of their awareness~\cite{smith2013choice, solove2012introduction}. To ensure user autonomy and resolve the dilemma of privacy decision-making, we recommend future research to explore ways to personalize the privacy defaults based on user preferences. AI-supported XR systems could automatically predict user preferences based on prior use and apply relevant settings as the new default that is more likely to fit with user expectations~\cite{huang2020amazon}. However, users should also be able to easily override the default settings when they want other options~\cite{sunstein2014nudge} or if they feel that the system misinterpreted their preferences. 

\subsubsection{Customizable Consent and Data Flow Tailored to XR}

Many participants desired more effective control and explicit consent for specific types of data collection through XR software configurations. While privacy notifications are essential for enhancing informed data collection, research suggests that users feel greater privacy violations when they cannot manage their data collection and use~\cite{patil2015interrupt,bemmann2022influence}. Their frustration on the absence of privacy choices may further lead to privacy resignation~\cite{colnago2020informing}. Since privacy controls in XR could move beyond screen-based paradigms, we suggest providing more natural and intuitive privacy controls tailored to 3D environments to make users feel more empowered. For example, using gestures like a ``thumbs up'' for consent may be more intuitive than interacting with a screen-based ``I agree'' button. Emerging controls through gestures and object-oriented interfaces provide valuable opportunities for XR designers to help users make informed privacy decisions and consent to their data use. Future work could explore the best suited interaction types in XR for privacy notifications and control mechanisms.

Currently, there is a lack of guidance to support XR system designers and privacy engineers when designing privacy notices. The traditional design dimensions for the web and mobile interfaces (i.e., timing, channel, modality, control~\cite{schaub2015design}) and the extended design space for IoT (i.e., type, functionality, timing, channel, modality)~\cite{feng2021design} need to be expanded further for XR environments. For example, the modality design dimension can extend beyond visual, auditory, and haptic to include motion, neurofeedback (e.g., emotion), and even user-defined olfaction (e.g., scent), and gustation (e.g., taste)~\cite{rakkolainen2021technologies} to convey users' privacy decisions to XR systems. The effectiveness of existing usability evaluation frameworks~\cite{habib2022evaluating} needs assessment, and new evaluation frameworks tailored to XR environments need to be developed.

\subsubsection{Enforcing and Developing Transparent XR Data Practices} 
Many participants reported strategies such as trusting and relying on the manufacturer to do what had been stated in their privacy policy, and understanding XR data practices by reading user agreements, privacy policies, or legal documentation. They also expressed the need to mandate data collection and that the data should primarily be used to serve user interests due to the fact that VR application data policies are often vague~\cite{adams2018ethics}, and most data are used for commercial purposes 
without user consent~\cite{trimananda2022ovrseen}. These problems highlight the need for transparent data practices associated with XR devices. 
Since many data types explored in this study are not used in most other types of consumer technology, current data privacy regulations are inadequate to address the privacy challenges posed by the proliferation of sensors and uncertainties in XR data practices~\cite{dick2021balancing}.
The immersive nature of XR makes it difficult to mitigate risks by applying existing privacy policies and practices for other types of digital media. Future research could help product makers design transparent controls in their products and help policymakers identify and develop new regulations around sensitive data types and mandate appropriate consent mechanisms and controls. When possible, data minimization practices should be implemented around personal and potentially sensitive data types, whether the data is required for the functionalities of the service or to enhance personalization and user experience. Our study provides some indication of the data types that users consider sensitive and suggests that particular attention should be paid to data types that are currently considered non-personally identifiable in non-XR technologies (e.g., behavioural identifier)~\cite{gugenheimer2022novel, miller2020personal, pfeuffer2019behavioural}). Our study also raises the need to design intuitive notices and consent around passive data collection and make use of personalized default settings to help people manage their privacy and data in immersive XR experiences.



\subsection{Limitations}
\label{subsec:Limitations}

We note several limitations of our study. 
Our sample does not reflect the general population distribution of North America and Europe, as we did not include non-users in our study. Our survey was conducted only in English, non-English speaking XR users are inevitably excluded.  
Despite these limitations, our results offer valuable insights into the perceptions of those with experience using XR. Further studies could consider expanding the scope to understand the perspectives of non-users, the non-English speaking XR users, and the difference between users and non-users.


Our survey-based study methodology relies on the participants' self-reported privacy concerns, preferences, and behaviour, which might not match their actual behaviour of using XR devices.
Additionally, our study can suffer from bias when the scenarios describe situations that the participants have never encountered, affecting the precision of their perceptions and concerns. 
Although we recognize that mentioning data collection could influence participants' privacy concerns, efforts were made to minimize its impact. For example, we promoted our study as research about XR user experience and explicitly avoided using words like ``privacy,'' ``risk,'' and ``concern'' in the recruitment materials and questionnaire. We also acknowledge that the factors we examined in our scenarios were not an exhaustive list. Currently, much information regarding XR data practices is unclear~\cite{XRSI2021Virtual}. We hope that our study can provide valuable insights for future discussions and encourage researchers to explore user privacy concerns in more complex situations, such as bystander involvement, multiple user scenarios, and other related factors.

\section{Conclusion}
\label{sec:Conclusion}

Our paper reported a scenario-based study on privacy concerns related to XR technologies. We surveyed 464 participants and identified their (lack of) awareness of XR data collection, their levels of privacy concerns, factors that influence their concerns, and coping strategies that mitigate their discomfort caused by XR data collection. Our results indicated that participants' comfort regarding XR data collection is directly driven by type of data that are collected, perceived data sensitivity, perceived realism of the scenarios, and perceived equity of data collection. Furthermore, many reported coping strategies similar to web, mobile and IoT scenarios. Based on our findings, we discussed the need to address users' lack of awareness and correct privacy misconceptions originated from previous interactions with non-XR devices, the design of privacy interfaces tailored to immersive XR technologies, and the need to develop and mandate transparent data practices. We hope that our work can provide insight for future user-centred XR privacy studies. XR technologies are still in early stages, which provides an opportunity to proactively develop solutions and countermeasures to address potential privacy issues. 
Additionally, we hope that our finding will encourage and facilitate future privacy research for other XR user groups, settings, and usage scenarios, 
such as bystanders and multi-user environments. 

\begin{acks}
The research was funded by the Games Institute Seed Grant from the University of Waterloo. L. Zhang-Kennedy (\#RGPIN-2022-03353) and L. Nacke (\#RGPIN-2023-03705) also acknowledge support from the Natural Sciences and Engineering Research Council of Canada (NSERC) Discovery Grants. Any opinions, findings, conclusions, or recommendations expressed in this material are those of the author(s) and do not necessarily reflect the views of the Games Institute, the University of Waterloo, or NSERC.
\end{acks}

\nocite{*}  
\bibliographystyle{ACM-Reference-Format}
\bibliography{98-References}

\newpage
\appendix
\clearpage
\section{Supplementary Materials --- Survey Questions}

\subsection{User Understanding of XR Technologies}
\label{app_subsec:xr_understanding}

For the introductory purpose, the participants were first presented with a drag and drop pair-matching exercise, where they were asked to match the following descriptions (from~\citet{XRSI2020Definition}) with corresponding technologies: VR, AR, MR. After this exercise, we further reinforced participants' understanding by providing them with the correct VR, AR, and MR definitions~\cite{XRSI2020Definition}:

\textbf{Extended Reality (XR)} is a fusion of all the realities --- including \textbf{Virtual Reality (VR)}, \textbf{Augmented Reality (AR)}, and \textbf{Mixed Reality (MR)} --- which consists of technology-mediated experiences enabled via a wide spectrum of hardware and software, including sensory interfaces, applications, and infrastructures. 

\begin{itemize}
    \item \textbf{VR}
    \begin{itemize}
        \item It is a fully immersive software-generated, artificial digital environment. 
        \item It is a simulation of three-dimensional images, experienced by users via special electronic equipment, such as a headset. 
        \item It places the experiencer in another environment entirely. Whether that environment has been generated by a computer or captured by video, it entirely occludes the experiencer's natural surroundings.
    \end{itemize}
    \item \textbf{AR} 
    \begin{itemize}
        \item It overlays digital-created content on top of the user’s real-world environment, viewed through a device (such as a smartphone) that incorporates real-time inputs to create an enhanced version of reality. 
        \item Digital and virtual objects (e.g., graphics, sounds) are superimposed on an existing environment to create an immersive experience.
    \end{itemize}
    \item \textbf{MR} 
    \begin{itemize}
        \item It seamlessly blends the user’s real-world environment with digitally-created content, where both environments can coexist and interact with each other. 
        \item The virtual objects behave in all aspects as if they are present in the real-world, e.g., they are occluded by physical objects, they sound as though they are in the same space as the user. As the user interacts with the real and virtual objects, the virtual objects will reflect the changes in the environment as would any real object in the same space.
    \end{itemize}
\end{itemize}

\subsection{User Awareness of XR Data Operation}

\begin{enumerate}[label= \textbf{Q\arabic*:}]
    \item You mentioned that you have experience using XR devices. What data do you think are being observed by the device? 
    \begin{itemize}
        \item Identifying in-app/in-world assets (e.g., personal virtual objects, personal avatars)
        \item Physical appearance (e.g., body dimensions)
        \item Environment information and dimensions (e.g., users’ surroundings, room layout, device/user position in relation to environment, user position in relation to device) 
        \item Physical movements and characteristics (e.g., head/hand/body/eye motion, orientation, position, gestures, posture, fitness information)
        \item Physiological data (e.g., pupil and cornea reflections, brainwaves, skin signals)
        \item Visual attention (e.g., eye gaze, area of interest, fixation, heatmaps, time to first fixation, time spent on a certain point, fixation sequences) 
        \item Mannerisms (e.g., gait, habitual movements) 
        \item Cognitive, emotional, personality estimates (e.g., cognitive load, stress, depression, excitement, identity traits, etc.) 
        \item Non-identifying virtual assets (e.g., in-app achievements)
    \end{itemize}
\end{enumerate}

\subsection{Scenarios and Questions}

Sample Scenario:
\begin{quote}
 \textit{You are at home / personal room, and your XR device (e.g., smart headset, touch controller, 3D projects) is keeping track of your \textbf{Identifying in-app / in-world assets (e.g., personal virtual objects, personal avatars)} when \textbf{you use the device}.} 
\end{quote}

Questions below are repeated with each Scenario.
\begin{enumerate}[resume,label= \textbf{Q\arabic*:}]
    \item ``I think the Scenario like this (will) happen...'' (each answered on 5-point Likert scales from ``1-extremely unlikely'' to ``5-extremely likely'')
    \begin{itemize}
        \item Today
        \item Within 2 years
        \item Within 10 years
    \end{itemize}
    \item How comfortable are you about the data collection described in the Scenario? (answered on a 5-point Likert scale from ``1-very uncomfortable'' to ``5-very comfortable'')
    \item Please indicate how sensitive you consider the data being collected in the Scenario. \textbf{Sensitive Information} is any information which could cause serious mental, physical, or financial harm if it's lost, misused, or shared without permission. (answered on a 5-point Likert scale from ``1-not sensitive'' to ``5-very sensitive'')
    \item (if ``1-very uncomfortable'' or ``2-uncomfortable'' in Q3) If you had no choice on data collection, what would you do (if any) to mitigate the discomfort raised? [open-ended question]
\end{enumerate}


\begin{landscape}
\raggedright
\vspace*{\fill}
\begin{table}[htb!]
\caption[position=top]{Codebook --- This codebook presents the codes and themes we synthesized from participants' responses to the open-ended question ``If you had no choice on data collection, what would you do (if any) to mitigate the discomfort raised?'' (Q5).}
\label{tab:codebook}
\resizebox{0.8\paperheight}{!}{%
\begin{tabular}{@{}lll@{}}
\toprule
\multicolumn{1}{l|}{\textbf{Theme}} & \multicolumn{1}{l|}{\textbf{Frequent codes}} & \textbf{Example responses} \\ \midrule
\multicolumn{1}{l|}{\multirow{4}{*}{Away from device}} & \multicolumn{1}{l|}{Avoid at all cost} & ``It can be very dangerous. Avoid at all cost.'' \\
\multicolumn{1}{l|}{} & \multicolumn{1}{l|}{Cease using equipment} &``Most likely the only way to reduce data collection would be to not use the device.''\\
\multicolumn{1}{l|}{} & \multicolumn{1}{l|}{Get rid of device} & ``I would get rid of device.'' ``I'd sell the VR headset.'' \\
\multicolumn{1}{l|}{} & \multicolumn{1}{l|}{Wouldn't buy the device} &``I'd not purchase device which is capable of this kind of data collection.'' \\ \midrule
\multicolumn{1}{l|}{\multirow{3}{*}{Benefit vs. Privacy}} & \multicolumn{1}{l|}{Accept the data collection if its beneficial to me} & ``If them obtaining that data benefited me in some way, improved my health for example.'' \\
\multicolumn{1}{l|}{} & \multicolumn{1}{l|}{Evaluate if the device worth the value} & ``ask myself is it worth it.If the reasons are strong enough I [will] convince myself to only focus on the price.'' \\
\multicolumn{1}{l|}{} & \multicolumn{1}{l|}{I should be paid if they use my data} & ``at least reward people for their own data.'' ``[I should] getting paid for the collection and use of my data.''\\ \midrule
\multicolumn{1}{l|}{\multirow{4}{*}{\begin{tabular}[c]{@{}c@{}}Company and System\\ Trust\end{tabular}}} & \multicolumn{1}{l|}{Bring concern to the company} & ``I'd try to raise concerns to such companies.'' \\
\multicolumn{1}{l|}{} & \multicolumn{1}{l|}{Choose the device/brands I trust} & ``choose the devices/brands I feel most comfortable sharing my data with.'' \\
\multicolumn{1}{l|}{} & \multicolumn{1}{l|}{Must comply with user privacy settings} & ``allow users to set privacy settings and actually comply with them.'' \\
\multicolumn{1}{l|}{} & \multicolumn{1}{l|}{Trust the system} & ``I'd trust a good system, so in the end I'd be comfortable sharing my data.'' \\ \midrule
\multicolumn{1}{l|}{\multirow{3}{*}{Consent and Permission}} & \multicolumn{1}{l|}{Be able to opt out} & ``I'd like to be able to opt out.'' \\
\multicolumn{1}{l|}{} & \multicolumn{1}{l|}{Data collection should with my permission} & ``Informing that it is doing this and when it is collecting data, how it is being used, and viewing it if necessary.'' ``I'd not partake at all if sensitive data was collected from me without my explicit permission.''\\
\multicolumn{1}{l|}{} & \multicolumn{1}{l|}{Access and control data usage post-collection} & ``Having access on how and from whom my personal data are used for.'' \\ \midrule
\multicolumn{1}{l|}{\multirow{4}{*}{\begin{tabular}[c]{@{}c@{}}Comprehend and control what happen to\\ my data\end{tabular}}} & \multicolumn{1}{l|}{Ensure data is used only for my interests} & ``show the users that data is only used for their interests and not with second intentions.'' ``data is not used for malicious actions.'' \\
\multicolumn{1}{l|}{} & \multicolumn{1}{l|}{Search option from the Internet} & ``search for third-party ways to cheat that device so it would process wrong data [not my personal data].'' \\
\multicolumn{1}{l|}{} & \multicolumn{1}{l|}{Seek privacy protection regulations} & ``I'd check the data privacy laws surrounding the product.'' ``make sure to have some sort of data collection protection contract.'' \\
\multicolumn{1}{l|}{} & \multicolumn{1}{l|}{Understand data operation} & ``try research at maximum about the data collection to prevent discomfort.'' \\ \midrule
\multicolumn{1}{l|}{\multirow{5}{*}{Ignore and Give up}} & \multicolumn{1}{l|}{Don't mind data collection but can't be 24/7} & ``I don't mind the data it collects, but every waking moment in my day would be uncomfortable.''\\
\multicolumn{1}{l|}{} & \multicolumn{1}{l|}{Everyone's being monitored} & ``everyone is being monitored anyway. I'm not special.'' ``Technology will overcome us.'' ``We have no privacy.'' \\
\multicolumn{1}{l|}{} & \multicolumn{1}{l|}{I can mitigate the discomfort} & ``I believe I have the ability to mitigate the discomfort.'' \\
\multicolumn{1}{l|}{} & \multicolumn{1}{l|}{Keep using device and ignore the data collection} & ``just not think about the data being collected.'' ``ignore the discomfort.'' \\
\multicolumn{1}{l|}{} & \multicolumn{1}{l|}{None} & ``None.'' ``Nothing can be done.''\\ \midrule
\multicolumn{1}{l|}{\multirow{9}{*}{Limit the collection}} & \multicolumn{1}{l|}{Disconnect / turnoff /uninstall when not use} & ``unplug the device when I'm not using it.'' ``disconnect the device from the Internet.''\\
\multicolumn{1}{l|}{} & \multicolumn{1}{l|}{Have separate devices for different activities} & ``having a separate device for activities that wouldn't mix with my main personal computer or devices.'' \\
\multicolumn{1}{l|}{} & \multicolumn{1}{l|}{Limit the device usage only when necessary} & ``Use the device a lot less.'' ``only use it for certain things and games.'' \\
\multicolumn{1}{l|}{} & \multicolumn{1}{l|}{Not register personal info} & ``I can lie about my digital self and not have a very personal experience.'' ``Not register the personal avatar on the device.'' \\
\multicolumn{1}{l|}{} & \multicolumn{1}{l|}{Physically block device sensors} & ``obscure the view of the sensor that the device is using to operate.'' ``cover the device when not in use.'' \\
\multicolumn{1}{l|}{} & \multicolumn{1}{l|}{Physically contain the tracking ability} & ``find a way to totally block any data that is sent.'' ``put the device in a room that is less personal.'' \\
\multicolumn{1}{l|}{} & \multicolumn{1}{l|}{Seek data protection apps} & ``I'd download some apps to stop it [data collection].'' \\
\multicolumn{1}{l|}{} & \multicolumn{1}{l|}{Seek data protection configuration options} & ``find if there's a setting that disables data collection for commercial purposes.'' ``there should be options to reduce/control the amount and type of data being collected.''\\
\multicolumn{1}{l|}{} & \multicolumn{1}{l|}{Self-censor privacy disclosure} & ``avoid my usual mannerism so that the machine can't learn them.'' ``try to not demonstrate my feelings to that device.'' \\ \midrule
\multicolumn{1}{l|}{\multirow{5}{*}{Privacy above anything}} & \multicolumn{1}{l|}{Be careful and not support it} & ``have more attention when i use it.'' ``I'd refuse to support it [data collection] in any way.'' ``I'd not agree for this kind of monitoring.''  \\
\multicolumn{1}{l|}{} & \multicolumn{1}{l|}{Concerned about impersonation} & ``it would be real easy to impersonate someone if those details were to be gathered.'' \\
\multicolumn{1}{l|}{} & \multicolumn{1}{l|}{Personalized advertisements and manipulation} & ``it will be used to sell me something in the future.'' ``it can be used for manipulate my behavior through the shopping, voting and important life decisions habits.'' \\
\multicolumn{1}{l|}{} & \multicolumn{1}{l|}{Ensure that I can't be identified from my data} & ``assure me somehow data is anoymized.'' ``something like a VPN would exist for those circumstances in the future, where my data is collected by third parties are unable to connect that data to your identity.'' \\
\multicolumn{1}{l|}{} & \multicolumn{1}{l|}{Switch to alternatives that respect privacy} & ``I'd try other companies who doing device who respect privacy.'' ``I'd search for alternative headsets without data collection.'' \\ \bottomrule
\end{tabular}%
}
\end{table}
\vspace*{\fill}
\end{landscape}


\begin{table*}[p]
\centering
\caption{Overview of Participants' demographics ($N=464$)}
\label{tab:demographic}
\resizebox{0.5\textwidth}{!}{%
\begin{tabular}{@{}ll@{}}
\toprule
\textbf{Gender} & n (\% of total) \\ \midrule
Man & 230 (50\%)  \\
Woman & 222 (48\%) \\
Non-binary / third gender & 11 (2\%)  \\
Prefer not to say & 1 (<1\%) \\ \midrule
\multicolumn{2}{l}{\textbf{Age}} \\ \midrule
18-19 & 12 (3\%) \\
20-29 & 313 (68\%) \\
30-39 & 88 (19\%) \\
40-49 & 39 (8\%) \\
50-56 & 12 (3\%) \\
Mean (SD) & 27.78 (7.95) \\ \midrule
\multicolumn{2}{l}{\textbf{Education level}} \\ \midrule
Less than a high school diploma & 4 (<1\%) \\
High school degree or   equivalent & 173 (37\%)  \\
Associate's degree (e.g.,   AA, AS) & 30 (7\%)  \\
Bachelor's degree (e.g., BA,   BS) & 162 (35\%) \\
Graduate or professional   degree (e.g., MA, MD, Ph.D.) & 94 (20\%)  \\
Prefer not to say & 1 (<1\%) \\ \midrule
\multicolumn{2}{l}{\textbf{Income}} \\ \midrule
Less than \$25,000 & 182 (39\%)  \\
\$25,000-\$49,999 & 151 (33\%) \\
\$50,000-\$99,999 & 86 (19\%) \\
\$100,000-\$199,999 & 19 (4\%) \\
More than \$200,000 & 5 (1\%) \\
I prefer not to respond & 21 (5\%) \\ \midrule
\multicolumn{2}{l}{\textbf{Country of residence}} \\ \midrule
Belgium & 5 (1\%) \\
Canada & 5 (1\%)\\
Czech Republic & 8 (2\%)\\
Greece & 25 (5\%)\\
Hungary & 20 (4\%)\\
Italy & 43 (9\%)\\
Netherlands & 5 (1\%)\\
Poland & 66 (14\%)\\
Portugal & 123 (27\%)\\
Slovenia & 7 (2\%)\\
Spain & 28 (6\%)\\
United Kingdom of Great Britain and Northern Ireland & 57 (12\%)\\
United States of America & 52 (11\%)\\ 
Other (9 countries)* & 20 (4\%)\\
\midrule
\multicolumn{2}{l}{\textbf{Experience with XR devices}  (\varname{experience\_years})} \\ \midrule
Less than 6 months & 185 (40\%)\\
6 months to 1 year & 103 (22\%)\\
1 year to 5 years & 165 (36\%)\\
5 to 10 years & 9 (2\%)\\
More than 10 years & 2 (<1\%)\\ \midrule
\multicolumn{2}{l}{\textbf{Type of devices used} (\varname{experience\_device})} \\ \midrule
Smart headsets, smart glasses & 396 (85\%)\\
Touch controllers, wired gloves, 3D mouse & 121 (26\%)\\
Handheld or Mobile devices   (e.g., handheld viewers, mobile phone, tablet) & 268 (58\%)\\
Projectors and display walls & 51 (11\%)\\ \midrule
\multicolumn{2}{l}{\textbf{Frequency of use} (\varname{experience\_frequency})} \\ \midrule
At least once a day & 23 (5\%)\\
At least once a week & 117 (25\%)\\
At least once a month & 121 (26\%)\\
At least once every three   months & 69 (15\%)\\
At least once every six months & 60 (13\%)\\
Once a year & 74 (16\%)\\ \midrule
\multicolumn{2}{l}{\textbf{Purpose of use} (\varname{experience\_purpose})} \\ \midrule
Arts (e.g., virtual gallery, design \& prototyping, virtual graffiti) & 79 (17\%)\\
Education or training (e.g.,   students, sports, military, medical procedures) & 80 (17\%)\\
Entertainment (e.g., gaming, socializing) & 437 (94\%)\\
Healthcare (e.g., mental health diagnoses and treatment) & 11 (2\%)\\
Workplace requirement (e.g., remote work, staff connection, workplace functionality) & 31 (7\%)\\
Other (i.e., research, curiosity, getting directions, fitness)  & 10 (2\%)\\ \midrule
\multicolumn{2}{l}{\textbf{Internet Users' Information Privacy Concerns (IUIPC)}} \\ \midrule
\multicolumn{2}{l}{\underline{\textit{\textbf{Control factor}}} (\varname{IUIPC\_awareness})} \\
Range (Min - Max) & (3 - 7) \\
Mean (SD) & 5.78 (0.92) \\
\multicolumn{2}{l}{\underline{\textit{\textbf{Awareness factor}}} (\varname{IUIPC\_control})} \\
Range (Min - Max) & (3 - 7) \\
Mean (SD) & 6.29 (0.74) \\
\underline{\textit{\textbf{Collection factor}}} (\varname{IUIPC\_collection}) &  \\
Range (Min - Max) & (1.25 - 7) \\
Mean (SD) & 5.69 (1.14) \\ \bottomrule
\multicolumn{2}{l}{\textit{\begin{tabular}[c]{@{}l@{}}\small Note. * ``Other'' includes 9 countries, each containing fewer than 5 participants: Austria, Estonia, Finland, France,\\
\small Germany, Ireland, Lativa,  Norway, Sweden. \end{tabular}}} \\ 
\end{tabular}%
}
\end{table*}






\begin{landscape}
\raggedright
\vspace*{\fill}
\begin{table}[htb!]
\caption[position=top]{Wilcoxon Signed-Rank Tests with Bonferroni adjustment~\cite{chen2017general} on participants' comfort level and perceived data sensitivity across data types. A same pair of scenarios is presented to $n\approx64$ participants.}
\resizebox{0.69\paperheight}{!}{%
\begin{tabular}{@{}rccllllllllll@{}}
\toprule
& \multicolumn{2}{c}{\textbf{Descriptive Statistics}} &  & \multicolumn{8}{c}{}\\ \cmidrule(r){2-3}
\textbf{\textit{data\_type}} & \textbf{Median (SD)} & \textbf{Range (Min-Max)} &  & \multicolumn{8}{c}{\textbf{Wilcoxon-signed rank test}} \\ \midrule
\textbf{\textit{comfort\_level}} &  &  &  & \rotatebox[origin=r]{270}{Physical appearance} & \rotatebox[origin=r]{270}{Identifying   in-app/in-world assets} & \rotatebox[origin=r]{270}{Environment information and dimensions} & \rotatebox[origin=r]{270}{Physical movements and characteristics} & \rotatebox[origin=r]{270}{Physiological data} & \rotatebox[origin=r]{270}{Visual attention} & \rotatebox[origin=r]{270}{Mannerisms} & \rotatebox[origin=r]{270}{Cognitive, emotional, and personality estimates} \\ 
Physical appearance & 3 (1.10) & (1 - 5) & & & & & & & & & \\
Identifying in-app/in-world assets & 3 (1.10)  & (1 - 5) & &\cellcolor{palered!50} \textit{W=248.5} &  &  &  &  &  &  &  \\
& & & &\cellcolor{palered!50} \textit{P=.002\sig{**}} &  &  &  &  &  &  &  \\
& & & &\cellcolor{palered!50} \textit{r=-0.39} &  &  &  &  &  &  &  \\ \cmidrule(r){5-12}
Environment information and dimensions & 3 (1.06) &(1 - 5) & &\textit{W=323.5} & \cellcolor{palered!50} \textit{W=804.5} &  &  &  &  &  &  \\
& & & &\textit{P<.15} & \cellcolor{palered!50} \textit{P=.001\sig{***}} &  &  &  &  &  &  \\
& & & &\textit{r=-0.17} & \cellcolor{palered!50} \textit{r=-0.46} &  &  &  &  &  &  \\\cmidrule(r){5-12}
Physical movements and characteristics & 3 (1.14) &(1 - 5) & &\cellcolor{palered!50} \textit{W=320.0,} & \textit{W=502} & \textit{P=.65} &  &  &  &  &  \\
& & & &\cellcolor{palered!50} \textit{P=.02\sig{*}} & \textit{P=.11} & \textit{P=.65} &  &  &  &  &  \\
& & & &\cellcolor{palered!50} \textit{r=-0.28} & \textit{r=-0.20} & \textit{r=-0.06} &  &  &  &  &  \\ \cmidrule(r){5-12}
Physiological data & 2 (1.03) &(1 - 5) & &\textit{W=612.5} & \cellcolor{palered!50} \textit{W=1473.5} & \cellcolor{palered!50} \textit{W=738.5} & \cellcolor{palered!50} \textit{W=838.5} &  &  &  &  \\
&&& &\textit{P=.81} & \cellcolor{palered!50} \textit{P<.001\sig{***}} & \cellcolor{palered!50} \textit{P=.003\sig{**}} & \cellcolor{palered!50} \textit{P<.001\sig{***}} &  &  &  &  \\
&&& &\textit{r=-0.03} & \cellcolor{palered!50} \textit{r=-0.48} & \cellcolor{palered!50} \textit{r=-0.37} & \cellcolor{palered!50} \textit{r=-0.46} &  &  &  &  \\ \cmidrule(r){5-12}
Visual attention & 3 (1.12)  & (1 - 5)& &\textit{W=395.5} & \cellcolor{palered!50} \textit{W=410.5} & \textit{W=465} & \textit{W=367} & \cellcolor{palered!50} \textit{W=181.5} &  &  &  \\
&&& &\textit{P=.64} & \cellcolor{palered!50} \textit{P=.048\sig{*}} & \textit{P=.46} & \textit{P=.40} & \cellcolor{palered!50} \textit{P=.04\sig{*}} &  &  &  \\
&&& &\textit{r=-0.06} & \cellcolor{palered!50} \textit{r=-0.25} & \textit{r=-0.10} & \textit{r=-0.10} & \cellcolor{palered!50} \textit{r=-0.28} &  &  &  \\ \cmidrule(r){5-12}
Mannerisms &  3 (1.07)  & (1 - 5)& &\textit{W=345.0} & \cellcolor{palered!50} \textit{W=827} & \textit{W=539.5} & \textit{W=574.5} & \textit{W=231.5} & \textit{W=584.5} &  &  \\
&&& &\textit{P=.62} & \cellcolor{palered!50} \textit{P<.001\sig{***}} & \textit{P=.25} & \textit{P=.12} & \textit{P=.06} & \textit{P=.29} &  &  \\
&&& &\textit{r=-0.07} & \cellcolor{palered!50} \textit{r=-0.49} & \textit{r=-0.14} & \textit{r=-0.18} & \textit{r=-0.24} & \textit{r=-0.13} &  &  \\ \cmidrule(r){5-12}
Cognitive, emotional, and personality estimates &  2 (1.14) &(1 - 5) & &\cellcolor{palered!50} \textit{W=1052.5} & \cellcolor{palered!50} \textit{W=1031.5} & \cellcolor{palered!50} \textit{W=751} & \cellcolor{palered!50} \textit{W=1002} & \cellcolor{palered!50} \textit{W=317} & \cellcolor{palered!50} \textit{W=988} & \cellcolor{palered!50} \textit{W=567.5} &  \\
&&& &\cellcolor{palered!50} \textit{P<.001\sig{***}} & \cellcolor{palered!50} \textit{P<.001\sig{***}} & \cellcolor{palered!50} \textit{P<.001\sig{***}} & \cellcolor{palered!50} \textit{P<.001\sig{***}} & \cellcolor{palered!50} \textit{P=.007\sig{**}} & \cellcolor{palered!50} \textit{P=.002\sig{**}} & \cellcolor{palered!50} \textit{P=.001\sig{**}} &  \\
&&& &\cellcolor{palered!50} \textit{r=-0.59} & \cellcolor{palered!50} \textit{r=-0.61} & \cellcolor{palered!50} \textit{r=-0.61} & \cellcolor{palered!50} \textit{r=-0.50} & \cellcolor{palered!50} \textit{r=-0.39} & \cellcolor{palered!50} \textit{r=-0.39} & \cellcolor{palered!50} \textit{r=-0.42} &  \\ \cmidrule(r){5-12}
Non-identifying virtual assets &  4 (1.01)  & (1 - 5)& &\cellcolor{palered!50} \textit{W=143.5} & \cellcolor{palered!50} \textit{W=189} & \cellcolor{palered!50} \textit{W=285} & \cellcolor{palered!50} \textit{W=252} & \cellcolor{palered!50} \textit{W=0} & \cellcolor{palered!50} \textit{W=181} & \cellcolor{palered!50} \textit{W=146.5} & \cellcolor{palered!50} \textit{W=27.5} \\
&&& &\cellcolor{palered!50} \textit{P<.001\sig{***}} & \cellcolor{palered!50} \textit{P=.001\sig{**}} & \cellcolor{palered!50} \textit{P<.001\sig{***}} & \cellcolor{palered!50} \textit{P=.005\sig{**}} & \cellcolor{palered!50} \textit{P<.001\sig{***}} & \cellcolor{palered!50} \textit{P<.001\sig{***}} & \cellcolor{palered!50} \textit{P<.001\sig{***}} & \cellcolor{palered!50} \textit{P<.001\sig{***}} \\
&&& &\cellcolor{palered!50} \textit{r=-0.58} & \cellcolor{palered!50} \textit{r=-0.40} & \cellcolor{palered!50} \textit{r=-0.53} & \cellcolor{palered!50} \textit{r=-0.35} & \cellcolor{palered!50} \textit{r=-0.80} & \cellcolor{palered!50} \textit{r=-0.57} & \cellcolor{palered!50} \textit{r=-0.62} & \cellcolor{palered!50} \textit{r=-0.78} \\ \bottomrule
\multicolumn{12}{l}{\begin{tabular}[c]{@{}l@{}} \textit{Note. W=Test Statistic, r=Effect size, Significance are displayed as: \sig{***} P<.001, \sig{**} P<.01, \sig{*} P<.05.} \textit{Table continued on next page.}\end{tabular}} \\ 
\end{tabular}%
}
\label{tab:wilcoxon-signed-rank-data-type}
\end{table}
\vspace*{\fill}
\end{landscape}

\begin{landscape}
\raggedright
\vspace*{\fill}
\begin{table}[htb!]
\caption*{Table 7 continued. Wilcoxon Signed-Rank Tests with Bonferroni adjustment~\cite{chen2017general} on participants' comfort level and perceived data sensitivity across data types. A same pair of scenarios is presented to $n\approx64$ participants.}
\resizebox{0.69\paperheight}{!}{%
\begin{tabular}{@{}rccllllllllll@{}}
\toprule
& \multicolumn{2}{c}{\textbf{Descriptive Statistics}} &  & \multicolumn{8}{c}{}\\ \cmidrule(r){2-3}
\textbf{\textit{data\_type}} & \textbf{Median (SD)} & \textbf{Range (Min-Max)} &  & \multicolumn{8}{c}{\textbf{Wilcoxon-signed rank test}} \\ \midrule
\textbf{\textit{data\_sensitivity}} &  &  &  & \rotatebox[origin=r]{270}{Physical appearance} & \rotatebox[origin=r]{270}{Identifying   in-app/in-world assets} & \rotatebox[origin=r]{270}{Environment information and dimensions} & \rotatebox[origin=r]{270}{Physical movements and characteristics} & \rotatebox[origin=r]{270}{Physiological data} & \rotatebox[origin=r]{270}{Visual attention} & \rotatebox[origin=r]{270}{Mannerisms} & \rotatebox[origin=r]{270}{Cognitive, emotional, and personality estimates} \\
Physical appearance & 4 (1.28) & (1 - 5) & & & & & & & & & \\
Identifying in-app/in-world assets & 2 (1.24)  & (1 - 5)& &\cellcolor{palered!50} \textit{W=782.5} &  &  &  &  &  &  &  \\
& & & &\cellcolor{palered!50} \textit{P=.008\sig{**}} &  &  &  &  &  &  &  \\
& & & &\cellcolor{palered!50} \textit{r=-0.34} &  &  &  &  &  &  &  \\ \cmidrule(r){5-12}
Environment information and dimensions & 3 (1.20)  &(1 - 5) & &\textit{W=469} & \cellcolor{palered!50} \textit{W=164.5} &  &  &  &  &  &  \\
& & & &\textit{P=.42} & \cellcolor{palered!50} \textit{P<.001\sig{***}} &  &  &  &  &  &  \\
& & & &\textit{r=-0.10} & \cellcolor{palered!50} \textit{r=-0.49} &  &  &  &  &  &  \\ \cmidrule(r){5-12}
Physical movements and characteristics & 3 (1.26)  & (1 - 5)& &\textit{W=594} & \cellcolor{palered!50} \textit{W=181} & \textit{W=322} &  &  &  &  &  \\
& & & &\textit{P=.23} & \cellcolor{palered!50} \textit{P<.001\sig{***}} & \textit{P=.67} &  &  &  &  &  \\
& & & &\textit{r=-0.15} & \cellcolor{palered!50} \textit{r=-0.49} & \textit{r=-0.06} &  &  &  &  &  \\ \cmidrule(r){5-12}
Physiological data &  4 (1.16) &(1 - 5) & &\textit{W=503.5} & \cellcolor{palered!50} \textit{W=234} & \cellcolor{palered!50} \textit{W=327.5} & \cellcolor{palered!50} \textit{W=151} &  &  &  &  \\
& & & &\textit{P=.68} & \cellcolor{palered!50} \textit{P<.001\sig{***}} & \cellcolor{palered!50} \textit{P=.006\sig{**}} & \cellcolor{palered!50} \textit{P<.001\sig{***}} &  &  &  &  \\
& & & &\textit{r=-0.05} & \cellcolor{palered!50} \textit{r=-0.58} & \cellcolor{palered!50} \textit{r=-0.34} & \cellcolor{palered!50} \textit{r=-0.46} &  &  &  &  \\ \cmidrule(r){5-12}
Visual attention &  3 (1.22) &(1 - 5) & &\textit{W=547.5} & \textit{W=276.5} & \textit{W=369} & \cellcolor{palered!50} \textit{W=526.5} & \cellcolor{palered!50} \textit{W=406} &  &  &  \\
& & & &\textit{P=.12} & \textit{P=.07} & \textit{P=.57} & \cellcolor{palered!50} \textit{P=.006\sig{**}} & \cellcolor{palered!50} \textit{P=.002\sig{**}} &  &  &  \\
& & & &\textit{r=-0.20} & \textit{r=-0.24} & \textit{r=-0.07} & \cellcolor{palered!50} \textit{r=-0.34} & \cellcolor{palered!50} \textit{r=-0.44} &  &  &  \\ \cmidrule(r){5-12}
Mannerisms &  3 (1.20) & (1 - 5)&  &\textit{W=411} & \cellcolor{palered!50} \textit{W=262} & \textit{W=695.5} & \textit{W=286} & \cellcolor{palered!50} \textit{W=742.5} & \textit{W=290} &  &  \\
& & & &\textit{P=.56} & \cellcolor{palered!50} \textit{P=.002\sig{**}} & \textit{P=.40} & \textit{P=.09} & \cellcolor{palered!50} \textit{P<.001\sig{***}} & \textit{P=.16} &  &  \\
& & & &\textit{r=-0.08} & \cellcolor{palered!50} \textit{r=-0.38} & \textit{r=-0.10} & \textit{r=-0.20} & \cellcolor{palered!50} \textit{r=-0.48} & \textit{r=-0.18} &  &  \\ \cmidrule(r){5-12}
Cognitive, emotional, and personality estimates & 4 (1.00)  &(1 - 5) & &\cellcolor{palered!50} \textit{W=17} & \cellcolor{palered!50} \textit{W=113} & \cellcolor{palered!50} \textit{W=88} & \cellcolor{palered!50} \textit{W=75} & \cellcolor{palered!50} \textit{W=53} & \cellcolor{palered!50} \textit{W=74.5} & \cellcolor{palered!50} \textit{W=104} &  \\
& & & &\cellcolor{palered!50} \textit{P<.001\sig{***}} & \cellcolor{palered!50} \textit{P<.001\sig{***}} & \cellcolor{palered!50} \textit{P<.001\sig{***}} & \cellcolor{palered!50} \textit{P<.001\sig{***}} & \cellcolor{palered!50} \textit{P<.001\sig{***}} & \cellcolor{palered!50} \textit{P<.001\sig{***}} & \cellcolor{palered!50} \textit{P<.001\sig{***}} &  \\
& & & &\cellcolor{palered!50} \textit{r=-0.73} & \cellcolor{palered!50} \textit{r=-0.67} & \cellcolor{palered!50} \textit{r=-0.61} & \cellcolor{palered!50} \textit{r=-0.71} & \cellcolor{palered!50} \textit{r=-0.61} & \cellcolor{palered!50} \textit{r=-0.69} & \cellcolor{palered!50} \textit{r=-0.58} &  \\ \cmidrule(r){5-12}
Non-identifying virtual assets & 2 (1.28)  & (1 - 5)& &\cellcolor{palered!50} \textit{W=1109} & \cellcolor{palered!50} \textit{W=525.5} & \cellcolor{palered!50} \textit{W=1723.5} & \cellcolor{palered!50} \textit{W=1012.5} & \cellcolor{palered!50} \textit{W=1469.5} & \cellcolor{palered!50} \textit{W=831} & \cellcolor{palered!50} \textit{W=934.5} & \cellcolor{palered!50} \textit{W=1263} \\ 
& & & &\cellcolor{palered!50} \textit{P<.001\sig{***}} & \cellcolor{palered!50} \textit{P<.001\sig{***}} & \cellcolor{palered!50} \textit{P<.001\sig{***}} & \cellcolor{palered!50} \textit{P<.001\sig{***}} & \cellcolor{palered!50} \textit{P<.001\sig{***}} & \cellcolor{palered!50} \textit{P<.001\sig{***}} & \cellcolor{palered!50} \textit{P<.001\sig{***}} & \cellcolor{palered!50} \textit{P<.001\sig{***}} \\ 
& & & &\cellcolor{palered!50} \textit{r=-0.62} & \cellcolor{palered!50} \textit{r=-0.49} & \cellcolor{palered!50} \textit{r=-0.56} & \cellcolor{palered!50} \textit{r=-0.61} & \cellcolor{palered!50} \textit{r=-0.79} & \cellcolor{palered!50} \textit{r=-0.64} & \cellcolor{palered!50} \textit{r=-0.52} & \cellcolor{palered!50} \textit{r=-0.81} \\ \bottomrule
\multicolumn{12}{l}{\begin{tabular}[c]{@{}l@{}} \textit{Note. W=Test Statistic, r=Effect size, Significance are displayed as: \sig{***} P<.001, \sig{**} P<.01, \sig{*} P<.05.} \textit{Table continued on next page.}\end{tabular}} \\ 
\end{tabular}%
}
\label{tab:wilcoxon-signed-rank-data-type-part2}
\end{table}
\vspace*{\fill}
\end{landscape}

\begin{landscape}
\raggedright
\vspace*{\fill}
\begin{table}[htb!]
\caption*{Table 7 continued. Wilcoxon Signed-Rank Tests with Bonferroni adjustment~\cite{chen2017general} on participants' comfort level and perceived data sensitivity across data types. A same pair of scenarios is presented to $n\approx64$ participants.}

\resizebox{0.69\paperheight}{!}{%
\begin{tabular}{@{}rccllllllllll@{}}
\toprule
& \multicolumn{2}{c}{\textbf{Descriptive Statistics}} &  & \multicolumn{8}{c}{}\\ \cmidrule(r){2-3}
\textbf{\textit{data\_type}} & \textbf{Median (SD)} & \textbf{Range (Min-Max)} &  & \multicolumn{8}{c}{\textbf{Wilcoxon-signed rank test}} \\ \midrule
\textbf{\textit{likely\_today}} &  &  &  & \rotatebox[origin=r]{270}{Physical appearance} & \rotatebox[origin=r]{270}{Identifying   in-app/in-world assets} & \rotatebox[origin=r]{270}{Environment information and dimensions} & \rotatebox[origin=r]{270}{Physical movements and characteristics} & \rotatebox[origin=r]{270}{Physiological data} & \rotatebox[origin=r]{270}{Visual attention} & \rotatebox[origin=r]{270}{Mannerisms} & \rotatebox[origin=r]{270}{Cognitive, emotional, and personality estimates} \\ 
Physical appearance & 3 (1.27) & (1 - 5) & & & & & & & & & \\
Identifying in-app/in-world assets &  4 (1.28) & (1 - 5) &  &\cellcolor{palered!50} \textit{W=226.5} &  &  &  &  &  &  &  \\
& & & &\cellcolor{palered!50} \textit{P<.001\sig{***}} &  &  &  &  &  &  &  \\
& & & &\cellcolor{palered!50} \textit{r=-0.46} &  &  &  &  &  &  &  \\ \cmidrule(r){5-12}
Environment information and dimensions & 4 (1.26)  & (1 - 5) &  &\cellcolor{palered!50} \textit{W=217} & \textit{W=598} &  &  &  &  &  &  \\
& & & &\cellcolor{palered!50} \textit{P=.009\sig{**}} & \textit{P=.06} &  &  &  &  &  &  \\
& & & &\cellcolor{palered!50} \textit{r=-0.31} & \textit{r=-0.23} &  &  &  &  &  &  \\ \cmidrule(r){5-12}
Physical movements and characteristics & 4 (1.23)  &(1 - 5)  &  & \cellcolor{palered!50} \textit{W=339.5} & \textit{W=683} & \textit{W=217} &  &  &  &  &  \\
& & & &\cellcolor{palered!50} \textit{P=.002\sig{**}} & \textit{P=.11} & \textit{P=.37} &  &  &  &  &  \\
& & & &\cellcolor{palered!50} \textit{r=-0.37} & \textit{r=-0.20} & \textit{r=-0.12} &  &  &  &  &  \\ \cmidrule(r){5-12}
Physiological data & 2 (1.17)  &(1 - 5) & &\textit{W=520.5} & \cellcolor{palered!50} \textit{W=1402} & \cellcolor{palered!50} \textit{W=846.5} & \cellcolor{palered!50} \textit{W=797} &  &  &  &  \\
& & & &\textit{P=.24} & \cellcolor{palered!50} \textit{P<.001\sig{***}} & \cellcolor{palered!50} \textit{P<.001\sig{***}} & \cellcolor{palered!50} \textit{P<.001\sig{***}} &  &  &  &  \\
& & & &\textit{r=-0.14} & \cellcolor{palered!50} \textit{r=-0.66} & \cellcolor{palered!50} \textit{r=-0.57} & \cellcolor{palered!50} \textit{r=-0.54} &  &  &  &  \\ \cmidrule(r){5-12}
Visual attention & 3 (1.25)  &(1 - 5)  & &\cellcolor{palered!50} \textit{W=226} & \textit{W=471} & \cellcolor{palered!50} \textit{W=690.5} & \textit{W=361} & \cellcolor{palered!50} \textit{W=62} &  &  &  \\
& & & &\cellcolor{palered!50} \textit{P=.03\sig{*}} & \textit{P=.14} & \cellcolor{palered!50} \textit{P=.02\sig{*}} & \textit{P=.27} & \cellcolor{palered!50} \textit{P<.001\sig{***}} &  &  &  \\
& & & &\cellcolor{palered!50} \textit{r=-0.28} & \textit{r=-0.19} & \cellcolor{palered!50} \textit{r=-0.30} & \textit{r=-0.14} & \cellcolor{palered!50} \textit{r=-0.51} &  &  &  \\ \cmidrule(r){5-12}
Mannerisms & 3 (1.20)  & (1 - 5) & &\textit{W=372} & \cellcolor{palered!50} \textit{W=985.5} & \cellcolor{palered!50} \textit{W=637} & \cellcolor{palered!50} \textit{W=656} & \cellcolor{palered!50} \textit{W=97.5} & \textit{W=487} &  &  \\
& & & &\textit{P=.35} & \cellcolor{palered!50} \textit{P<.001\sig{***}} & \cellcolor{palered!50} \textit{P=.007\sig{**}} & \cellcolor{palered!50} \textit{P=.001\sig{**}} & \cellcolor{palered!50} \textit{P<.001\sig{***}} & \textit{P=.46} &  &  \\
& & & &\textit{r=-0.12} & \cellcolor{palered!50} \textit{r=-0.56} & \cellcolor{palered!50} \textit{r=-0.33} & \cellcolor{palered!50} \textit{r=-0.40} & \cellcolor{palered!50} \textit{r=-0.54} & \textit{r=-0.09} &  &  \\ \cmidrule(r){5-12}
Cognitive, emotional, and personality estimates &  2 (1.08) & (1 - 5)  &  &\cellcolor{palered!50} \textit{W=1016.5} & \cellcolor{palered!50} \textit{W=1387} & \cellcolor{palered!50} \textit{W=982.5} & \cellcolor{palered!50} \textit{W=968} & \textit{W=259} & \cellcolor{palered!50} \textit{W=1006} & \cellcolor{palered!50} \textit{W=811} &  \\
& & & &\cellcolor{palered!50} \textit{P<.001\sig{***}} & \cellcolor{palered!50} \textit{P<.001\sig{***}} & \cellcolor{palered!50} \textit{P<.001\sig{***}} & \cellcolor{palered!50} \textit{P<.001\sig{***}} & \textit{P=.09} & \cellcolor{palered!50} \textit{P<.001\sig{***}} & \cellcolor{palered!50} \textit{P<.001\sig{***}} &  \\
& & & &\cellcolor{palered!50} \textit{r=-0.59} & \cellcolor{palered!50} \textit{r=-0.67} & \cellcolor{palered!50} \textit{r=-0.70} & \cellcolor{palered!50} \textit{r=-0.70} & \textit{r=-0.24} & \cellcolor{palered!50} \textit{r=-0.54} & \cellcolor{palered!50} \textit{r=-0.53} &  \\ \cmidrule(r){5-12}
Non-identifying virtual assets &  4 (1.30)  & (1 - 5) & &\cellcolor{palered!50} \textit{W=108} & \textit{W=225} & \cellcolor{palered!50} \textit{W=485.5} & \cellcolor{palered!50} \textit{W=195.5} & \cellcolor{palered!50} \textit{W=10} & \cellcolor{palered!50} \textit{W=212} & \cellcolor{palered!50} \textit{W=111.5} & \cellcolor{palered!50} \textit{W=61} \\
& & & &\cellcolor{palered!50} \textit{P<.001\sig{***}} & \textit{P=.09} & \cellcolor{palered!50} \textit{P=.03\sig{*}} & \cellcolor{palered!50} \textit{P=.02\sig{*}} & \cellcolor{palered!50} \textit{P<.001\sig{***}} & \cellcolor{palered!50} \textit{P=.001\sig{**}} & \cellcolor{palered!50} \textit{P<.001\sig{***}} & \cellcolor{palered!50} \textit{P<.001\sig{***}} \\
& & & &\cellcolor{palered!50} \textit{r=-0.64} & \textit{r=-0.21} & \cellcolor{palered!50} \textit{r=-0.25} & \cellcolor{palered!50} \textit{r=-0.30} & \cellcolor{palered!50} \textit{r=-0.77} & \cellcolor{palered!50} \textit{r=-0.42} & \cellcolor{palered!50} \textit{r=-0.57} & \cellcolor{palered!50} \textit{r=-0.74} \\ \bottomrule
\multicolumn{12}{l}{\begin{tabular}[c]{@{}l@{}} \textit{Note. W=Test Statistic, r=Effect size, Significance are displayed as: \sig{***} P<.001, \sig{**} P<.01, \sig{*} P<.05.} \end{tabular}} \\ 
\end{tabular}%
}
\label{tab:wilcoxon-signed-rank-data-type-part3}
\end{table}
\vspace*{\fill}
\end{landscape}

\clearpage
\begin{table*}[!ht]
\centering
\caption{Wilcoxon Signed-Rank Tests on participants' comfort level across device statues ($N=464$).}
\resizebox{0.8\textwidth}{!}{%
\begin{tabular}{@{}lllll@{}}
\toprule
\multicolumn{1}{l}{\textbf{\textit{device\_status}}} & \multicolumn{2}{c}{\textbf{Descriptive Statistics}} &  & \\ \cmidrule(r){2-3}  
 & \textbf{Median (SD)} & \textbf{Range (Min-Max)} &  &  \multicolumn{1}{c}{\textbf{Wilcoxon Signed-Rank Test}} \\ \midrule
\multicolumn{1}{r}{\textbf{\textit{comfort\_level}}} & &  &  & In-use \\ 
In-use &  3 (1.10) & (1 - 5) &  & \\
Running in the background & 3 (1.20) & (1 - 5) &  & \cellcolor{palered!50}\textit{W=38588, P<.001\sig{***}, r=-0.23}  \\ \midrule
\multicolumn{1}{r}{\textbf{\textit{data\_sensitivity}}} &  &  &  & In-use \\ 
In-use & 3 (1.12) & (1 - 5) &  &  \\
Running in the background & 3 (1.10) & (1 - 5) &  & \cellcolor{palered!50}\textit{W=24216, P=.005\sig{**}, r=-0.14}  \\ \midrule
\multicolumn{1}{r}{\textbf{\textit{likely\_today}}}  &  &  &  & In-use \\ 
In-use & 3 (1.14) & (1 - 5) &  &  \\
Running in the background & 3 (1.11) & (1 - 5) &  & \cellcolor{palered!50}\textit{W=38028, P<.001\sig{***}, r=-0.24}  \\ \bottomrule
\multicolumn{5}{l}{\begin{tabular}[c]{@{}l@{}} \textit{Note. W=Test Statistic, r=Effect size, Significance are displayed as:\sig{***} P<.001, \sig{**} P<.01, \sig{*} P<.05.} \end{tabular}} \\ 
\end{tabular}%
}
\label{tab:wilcoxon-signed-rank-device-status}
\end{table*}
\clearpage

\end{document}